\documentclass[twocolumn,prc,showpacs,showkeys]{revtex4}
\usepackage{amsfonts}
\usepackage{amsmath}
\usepackage{amssymb}
\usepackage{graphicx}
\usepackage{rotating}

\providecommand{\U}[1]{\protect\rule{.1in}{.1in}}
\providecommand{\U}[1]{\protect\rule{.1in}{.1in}}
\providecommand{\U}[1]{\protect\rule{.1in}{.1in}}

\begin{document}

\title{Charged particle assisted nuclear reactions in solid state
environment: \\
renaissance of low energy nuclear physics\\
}
\author{P\'{e}ter K\'{a}lm\'{a}n\footnote{%
retired, e-mail: kalmanpeter3@gmail.com}}
\author{Tam\'{a}s Keszthelyi\thanks{%
retired, e-mail: khelyi@phy.bme.hu}}
\affiliation{Budapest University of Technology and Economics, Institute of Physics,
Budafoki \'{u}t 8. F., H-1521 Budapest, Hungary\ }

\begin{abstract}
The features of electron assisted neutron exchange processes in crystalline
solids are surveyed. It is found that, contrary to expectations, the cross
section of these processes may reach an observable magnitude even in the
very low energy case because of the extremely huge increment caused by the
Coulomb factor of the electron assisted processes and by the effect of the
crystal-lattice. The features of electron assisted heavy charged particle
exchange processes, electron assisted nuclear capture processes and heavy
charged particle assisted nuclear processes are also overviewed.
Experimental observations, which may be related to our theoretical findings,
are dealt with. A possible explanation of observations by Fleischmann and
Pons is presented. The possibility of the phenomenon of nuclear
transmutation is qualitatively explained with the aid of usual and charged
particle assisted reactions. The electron assisted neutron exchange
processes in pure $Ni$, $Pd$ and $Li-Ni$ composite systems (in the
Rossi-type E-Cat) are analyzed and it is concluded that the electron
assisted neutron exchange reactions in pure $Ni$ and $Li-Ni $ composite
systems may be responsible for recent experimental observations.
\end{abstract}

\startpage{1}
\endpage{}
\maketitle

\section{Introduction}

Since the "cold fusion" publication by Fleischmann and Pons in 1989 \cite%
{FP1} a new field of experimental physics has emerged. Although even the
possibility of the phenomenon of nuclear fusion at low energies is in doubt
in mainstream physics, the quest for low-energy nuclear reactions (LENR)
flourished and hundreds of publications (mostly experimental) have been
devoted to various aspects of the problem. (For the summary of experimental
observations, the theoretical efforts, and background events see e.g. \cite%
{Krivit}, \cite{Storms2}.) The main reasons for revulsion against the topic
according to standard nuclear physics have been: (a) due to the Coulomb
repulsion no nuclear reaction should take place at energies corresponding to
room temperature, (b) the observed extra heat attributed to nuclear
reactions is not accompanied by the nuclear end products expected from hot
fusion experiences, (c) traces of nuclear transmutations were also observed,
that considering the repulsive Coulomb interaction is an even more
inexplicable fact at these energies.

Motivated by the observations in the above field we search for physical
phenomena that may have modifying effect on nuclear reactions in solid state
environment. Earlier we theoretically found \cite{kk2}, \cite{kk0} that if
the reaction $p+d\rightarrow $ $^{3}He$ takes place in solid material then
the nuclear energy is mostly taken away by an electron of the environment
instead of the emission of a $\gamma $ photon, a result that calls the
attention to the possible role of electrons. Concerning the assistance of
the electrons and other charged constituents of the solid, a family of
electron assisted nuclear reactions, especially the electron assisted
neutron exchange process, furthermore the electron assisted nuclear capture
process and the heavy charged particle assisted nuclear processes were
discussed mostly in crystalline solid state (particularly in metal)
environment \cite{kk1}, \cite{kk3}. The aim of this paper is to summarize
our theoretical findings and on this basis to explain some experimental
observations.

We adopt the approach standard in nuclear physics when describing the cross
section of nuclear reactions. Accordingly, heavy, charged particles $j$ and $%
k$ of like positive charge of charge numbers $z_{j}$ and $z_{k}$ need
considerable amount of relative kinetic energy $E$ determined by the height
of the Coulomb barrier in order to let the probability of their nuclear
interaction have significant value. The cross section of such a process can
be\ derived applying the Coulomb solution $\varphi (\mathbf{r})$, 
\begin{equation}
\varphi (\mathbf{r})=e^{i\mathbf{k}\cdot \mathbf{r}}f(\mathbf{k,r})/\sqrt{V},
\label{Cb1}
\end{equation}%
which is the wave function of a free particle of charge number $z_{j}$ in a
repulsive Coulomb field of charge number $z_{k}$ \cite{Alder}, in the
description of relative motion of projectile and target. In $\left( \ref{Cb1}%
\right) $ $V$ denotes the volume of normalization, $\mathbf{r}$ is the
relative coordinate of the two particles, $\mathbf{k}$ is the wave number
vector in their relative motion and 
\begin{equation}
f(\mathbf{k},\mathbf{r})=e^{-\pi \eta _{jk}/2}\Gamma (1+i\eta
_{jk})_{1}F_{1}(-i\eta _{jk},1;i[kr-\mathbf{k}\cdot \mathbf{r}]),
\label{Hyperg}
\end{equation}%
where $_{1}F_{1}$ is the confluent hypergeometric function and $\Gamma $ is
the Gamma function. Since $\varphi (\mathbf{r})\sim e^{-\pi \eta
_{jk}/2}\Gamma (1+i\eta _{jk})$, the cross section of the process is
proportional to 
\begin{equation}
\left\vert e^{-\pi \eta _{jk}/2}\Gamma (1+i\eta _{jk})\right\vert ^{2}=\frac{%
2\pi \eta _{jk}\left( E\right) }{\exp \left[ 2\pi \eta _{jk}\left( E\right) %
\right] -1}=F_{jk}(E),  \label{Fjk}
\end{equation}%
the so-called Coulomb factor. Here 
\begin{equation}
\eta _{jk}\left( E\right) =z_{j}z_{k}\alpha _{f}\sqrt{a_{jk}\frac{m_{0}c^{2}%
}{2E}}  \label{etajk}
\end{equation}%
is the Sommerfeld parameter in the case of colliding particles of mass
numbers $A_{j}$, $A_{k}$ and rest masses $m_{j}=A_{j}m_{0}$, $%
m_{k}=A_{k}m_{0}$. $m_{0}c^{2}=931.494$ $MeV$ is the atomic energy unit, $%
\alpha _{f}$ is the fine structure constant and $E$ \ is taken in the center
of mass $\left( CM\right) $ coordinate system.%
\begin{equation}
a_{jk}=\frac{A_{j}A_{k}}{A_{j}+A_{k}}  \label{ajk}
\end{equation}%
is the reduced mass number of particles $j$ and $k$ of mass numbers $A_{j}$
and $A_{k}$. Thus the rate of the nuclear reaction of heavy, charged
particles of like positive charge becomes very small at low energies as a
consequence of $F_{jk}(E)$ being very small.

In the processes investigated the Coulomb and the strong interactions play
crucial role. The interaction Hamiltonian $H_{I}$ comprises the Coulomb
interaction potential $V_{Cb}$ with the charged constituents of surroundings
(solid) and the interaction potential $V_{St}$ of the strong interaction: 
\begin{equation}
H_{I}=V_{Cb}+V_{St}.  \label{HI}
\end{equation}%
(The Coulomb interaction between the charged participants of the nuclear
reaction is taken into account using $\left( \ref{Cb1}\right) $.) Therefore
the charged particle assisted nuclear reactions are at least second order in
terms of standard perturbation calculation. According to $\left( \ref{HI}%
\right) $, the lowest order of S-matrix element of a charged particle
assisted nuclear reaction has two terms which can be visualized with the aid
of two graphs. However, the contribution by the term, in which $V_{St}$
according to chronological order precedes $V_{Cb}$, is negligible because of
the smallness of the Coulomb factor the root square of which is appearing in
the matrix element of $V_{St}$ in this case. (In the following we only
depicts the graph of the dominant term.)

When describing the effect of the Coulomb interaction between the nucleus of
charge number $Z$ and a slow electron one can also use Coulomb function,
consequently, the cross section of the process to be investigated is
proportional to 
\begin{equation}
F_{e}(E)=\frac{2\pi \eta _{e}\left( E\right) }{\exp \left[ 2\pi \eta
_{e}\left( E\right) \right] -1},  \label{FeE}
\end{equation}%
but with 
\begin{equation}
\eta _{e}=-Z\alpha _{f}\sqrt{\frac{m_{e}c^{2}}{2E}}.  \label{eatae}
\end{equation}%
Here $m_{e}$ is the rest mass of the electron. In the case of low (less than 
$1$ $keV$) kinetic energy of the electron $F_{e}(E)$ reads approximately as $%
F_{e}(E)=\left\vert 2\pi \eta _{e}\left( E\right) \right\vert >1$.

For instance, the cross section of electron assisted neutron exchange
process (as it will be discussed later, and the graph of which is depicted
in Fig. 1) is proportional to $F_{e}(E)$ only (instead of $F_{jk}(E)$) since
the neutron takes part in strong interaction and so the corresponding matrix
element does not contain Coulomb factor. The increment in the cross section
due to changing $F_{jk}(E)$ for $F_{e}(E)$ in the case of electron assisted
neutron exchange process can be characterized by the ratio $%
F_{e}(E)/F_{jk}(E)$ which is an extremely large number. The cross section of
electron assisted neutron exchange process has a further (about a factor $%
10^{22}$) increase due to the presence of the lattice since the cross
section is also proportional to $1/v_{c}$. Here $v_{c}\sim d^{3}$ is the
volume of the elementary cell of the solid with $d$ the lattice parameter of
order of magnitude of $10^{-8}$ $cm$. The extremely huge increment in the
Coulomb factor increased further by the effect of the lattice makes it
possible that the cross section of electron assisted neutron exchange
process may reach an observable magnitude even in the very low energy case.
Thus it can be concluded that the actual Coulomb factors are the clue to the
charged particles assisted nuclear reactions and therefore we focus our
attention to them especially concerning the Coulomb factors of heavy charged
particles.

It is worth mentioning, that usual nuclear experiments, in which nuclear
reactions of heavy charged particles are investigated, are usually devised
taking into account the hindering effect of Coulomb repulsion. Consequently,
the beam energy is taken to be appropriately high to reach the energy domain
where the cross section of the processes becomes appropriately large.\
Therefore in an ordinary nuclear experiment the role of charged particle
assisted reactions is not essential.

\section{Applied method presented in electron assisted neutron exchange
process}

Recognizing the possibility and advantage of the assistance of electrons in
LENR we consider first the electron assisted neutron exchange process,
namely the

\begin{equation}
e+\text{ }_{Z_{1}}^{A_{1}}X+\text{ }_{Z_{2}}^{A_{2}}Y\rightarrow e^{\prime }+%
\text{ }_{Z_{1}}^{A_{1}-1}X+\text{ }_{Z_{2}}^{A_{2}+1}Y+\Delta
\label{exchange}
\end{equation}%
reaction \cite{kk3}\ (see Fig.1). Here $e$ and $e^{\prime }$ denote electron
and $\Delta $ is the energy of the reaction, i.e. the difference between the
rest energies of initial $\left( _{Z_{1}}^{A_{1}}X+_{Z_{2}}^{A_{2}}Y\right) $
and final $\left( _{Z_{1}}^{A_{1}-1}X+\text{ }_{Z_{2}}^{A_{2}+1}Y\right) $
states.

In $\left( \ref{exchange}\right) $ the electron (particle $1$) Coulomb
interacts with the nucleus $_{Z_{1}}^{A_{1}}X$ (particle $2$). A scattered
electron (particle $1^{\prime }$), the intermediate neutron (particle $3$)
and the nucleus $_{Z_{1}}^{A_{1}-1}X$ (particle $2^{\prime }$) are created
due to this interaction. The intermediate neutron (particle $3$) is captured
due to the strong interaction by the nucleus $_{Z_{2}}^{A_{2}}Y$ (particle $%
4 $) forming the nucleus $_{Z_{2}}^{A_{2}+1}Y$ (particle $5$) in this
manner. All told, in $\left( \ref{exchange}\right) $ the nucleus $%
_{Z_{1}}^{A_{1}}X$ (particle $2$) looses a neutron which is taken up by the
nucleus $_{Z_{2}}^{A_{2}}Y$ (particle $4$). The process is energetically
forbidden if $\Delta <0$. It was found, as it will be seen later, that the
electron takes away negligible energy. In this process the Coulomb factor of
electrons arises only since the particle, which is exchanged, is a neutron.

\begin{figure}[tbp]
\resizebox{6.0cm}{!}{\includegraphics{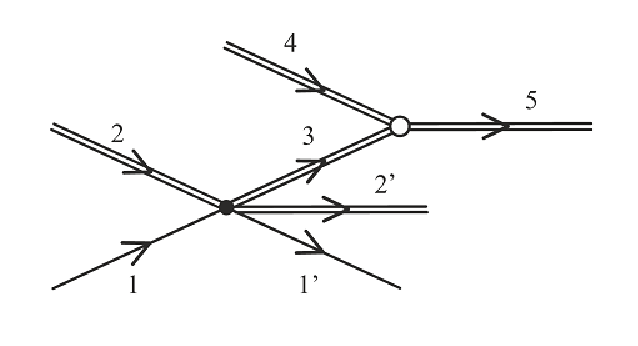}}
\caption{The graph of electron assisted neutron exchange process. Particle 1
(and 1') is an electron, particle 2 is a nucleus which looses a neutron and
becomes particle 2'. Particle 3 is an intermediate neutron. Particle 4 is
the nucleus which absorbs the neutron and becomes particle 5. The filled dot
denotes Coulomb-interaction and the open circle denotes nuclear (strong)
interaction. }
\label{figure1}
\end{figure}

The physical background to the virtual neutron stripping due to the Coulomb
interaction is worth mentioning. The attractive Coulomb interaction acts
between the $Z_{1}$ protons and the electron. The neutrons do not feel
Coulomb interaction. So one can say that in fact the nucleus $%
_{Z_{1}}^{A_{1}-1}X$ is stripped of the neutron due to the Coulomb
attraction.

As an example we take $Ni$ and $Pd$ as target material. It is thought that
the metal ($Ni$ or $Pd$) is irradiated with slow, free electrons. In this
case reaction $\left( \ref{exchange}\right) $ reads as%
\begin{equation}
e+\text{ }_{Z}^{A_{1}}X+\text{ }_{Z}^{A_{2}}X\rightarrow e^{\prime }+\text{ }%
_{Z}^{A_{1}-1}X+\text{ }_{Z}^{A_{2}+1}X+\Delta  \label{exchange metal}
\end{equation}%
with $Z=Z_{1}=Z_{2}$.

Now we demonstrate our calculation. Let us take a solid (in our case a
metal) which is irradiated by a monoenergetic beam of slow, free electrons.
The corresponding sub-system Hamiltonians are $H_{solid}$ and $H_{e}$. It is
supposed that their eigenvalue problems are solved, and the complete set of
the eigenvectors of the two independent systems are known. The interaction
between them is the Coulomb interaction of potential $V^{Cb}\left( \mathbf{x}%
\right) $ and the other interaction that is taken into account between the
nucleons of the solid is the strong interaction potential $V^{St}\left( 
\mathbf{x}\right) $. In the second order process investigated an electron
takes part in a Coulomb scattering with an atomic nucleus of the solid. In
the intermediate state a \ virtual free neutron $n$ is created which is
captured due to the strong interaction with some other nucleus of the solid.
The reaction energy $\Delta $ is shared between the quasi-free final
electron and the two final nuclei which take part in the process. Since the
aim of this paper is to show the fundamentals of the main effect, the
simplest description is chosen.

The electron of charge $-e$ and the nucleus $_{Z}^{A_{1}}X$ of charge $Ze$
take part in Coulomb-interaction. We use a screened Coulomb potential of the
form%
\begin{equation}
V^{Cb}\left( \mathbf{x}\right) =-\frac{Ze^{2}}{2\pi ^{2}}\int \frac{1}{%
q^{2}+\lambda ^{2}}\exp \left( i\mathbf{q}\cdot \mathbf{x}\right) d\mathbf{q}
\label{Vcb1}
\end{equation}%
with screening parameter $\lambda $ and coupling strength $e^{2}=\alpha
_{f}\hbar c$. For the strong interaction the interaction potential 
\begin{equation}
V^{St}\left( \mathbf{x}\right) =-f\frac{\exp \left( -s\left\vert \mathbf{x}%
\right\vert \right) }{\left\vert \mathbf{x}\right\vert }  \label{VSt1}
\end{equation}%
is applied, where the strong coupling strength $f=0.08\hbar c$ \cite{Bjorken}
and $1/s$ is the range of the strong interaction. ($\hbar $ is the reduced
Planck constant, $c$ is the velocity of light and $e$ is the elementary
charge.)

According to the standard perturbation theory of quantum mechanics the
transition probability per unit time $\left( W_{fi}\right) $ of this second
order process can be written as%
\begin{equation}
W_{fi}=\frac{2\pi }{\hbar }\sum_{f}\left\vert T_{fi}\right\vert ^{2}\delta
(E_{f}-E_{i}-\Delta )  \label{Wfie}
\end{equation}%
with 
\begin{equation}
T_{fi}=\sum_{\mu }\frac{V_{f\mu }^{St}V_{\mu i}^{Cb}}{\Delta E_{\mu i}}.
\label{Tif}
\end{equation}%
Here $V_{\mu i}^{Cb}$ \ is the matrix element of the Coulomb potential
between the initial and intermediate states and $V_{f\mu }^{St}$ is the
matrix element of the potential of the strong interaction between the
intermediate and final states, furthermore%
\begin{equation}
\Delta E_{\mu i}=E_{\mu }-E_{i}-\Delta _{i\mu }.  \label{DeltaEmui}
\end{equation}%
$E_{i}$, $E_{\mu }$ and $E_{f}$ are the kinetic energies in the initial,
intermediate and final states, respectively, $\Delta $ is the reaction
energy,~and $\Delta _{i\mu }$ is the difference between the rest energies of
the initial $\left( _{Z}^{A_{1}}X\right) $ and intermediate $\left(
_{Z}^{A_{1}-1}X\text{ and }n\right) $ states.%
\begin{equation}
\Delta =\Delta _{-}+\Delta _{+},\text{ }\Delta _{i\mu }=\Delta _{-}-\Delta
_{n}  \label{Delta}
\end{equation}%
with 
\begin{equation}
\Delta _{-}=\Delta _{A_{1}}-\Delta _{A_{1}-1}\text{ and }\Delta _{+}=\Delta
_{A_{2}}-\Delta _{A_{2}+1}.  \label{Delta-}
\end{equation}%
$\Delta _{A_{1}}$, $\Delta _{A_{1}-1}$, $\Delta _{A_{2}}$, $\Delta
_{A_{2}+1} $ and $\Delta _{n}$\ are the energy excesses of the neutral atoms
of mass numbers $A_{1}$, $A_{1}-1$, $A_{2}$, $A_{2}+1$ and the neutron,
respectively. \cite{Shir}. The sum of initial kinetic energies $\left(
E_{i}\right) $ is neglected in the energy Dirac-delta $\delta
(E_{f}-E_{i}-\Delta )$ and $\Delta E_{\mu i}$ further on.

Now for the sake of simplicity we reindex the particles. Particle indexed
with $e$ is the electron, particle indexed with $1$ is initially the nucleus 
$_{Z}^{A_{1}}X$ (particle 2 in Fig. 1) and finally $_{Z}^{A_{1}-1}X$
(particle 2' in Fig. 1), particle indexed with $2$ is initially the nucleus $%
_{Z}^{A_{2}}X$ (particle 4 in Fig. 1) and finally $_{Z}^{A_{2}+1}X$
(particle 5 in Fig. 1). 
\begin{equation}
E_{f}=E_{fe}\left( \mathbf{k}_{fe}\right) +E_{f1}\left( \mathbf{k}%
_{1}\right) +E_{f2}\left( \mathbf{k}_{2}\right) ,  \label{Ef}
\end{equation}

\begin{equation}
E_{\mu }=E_{fe}\left( \mathbf{k}_{fe}\right) +E_{\mu 1}\left( \mathbf{k}%
_{1}\right) +E_{n}\left( \mathbf{k}_{n}\right) ,  \label{Em}
\end{equation}%
where 
\begin{equation}
E_{fj}\left( \mathbf{k}_{j}\right) =\frac{\hbar ^{2}\mathbf{k}_{j}^{2}}{%
2m_{j}}  \label{Efj}
\end{equation}%
is the kinetic energy, $\mathbf{k}_{fj}\equiv \mathbf{k}_{j}$ is the wave
vector and $m_{j}$ is the rest mass of particle indexed with $j$ in the
final state $\left( j=1,2\right) $. 
\begin{equation}
E_{n}\left( \mathbf{k}_{n}\right) =\frac{\hbar ^{2}\mathbf{k}_{n}^{2}}{2m_{n}%
}  \label{En}
\end{equation}%
is the kinetic energy, $\mathbf{k}_{n}$ is the wave vector in the
intermediate state and $m_{n}$ is the rest mass of the neutron. $E_{\mu
1}\left( \mathbf{k}_{1}\right) $ is the kinetic energy of the first particle
in the intermediate state, and $E_{\mu 1}\left( \mathbf{k}_{1}\right)
=E_{f1}\left( \mathbf{k}_{1}\right) $. The kinetic energy of the electron in
the initial and final state 
\begin{equation}
E_{ie}=\frac{\hbar ^{2}\mathbf{k}_{ie}^{2}}{2m_{e}}\text{ and }E_{fe}=\frac{%
\hbar ^{2}\mathbf{k}_{fe}^{2}}{2m_{e}}  \label{E1f}
\end{equation}%
with $\mathbf{k}_{ie}$ and $\mathbf{k}_{fe}$ denoting the wave vector of the
electron in the initial and final state. The initial wave vectors $\mathbf{k}%
_{i1}$ and $\mathbf{k}_{i2}$ of particles $1$ and $2$ are neglected. The
initial, intermediate and final states are determined in Appendix A., the $%
V_{\mu i}^{Cb}$, $V_{f\mu }^{St}$ matrix-elements are calculated in Appendix
B. and the transition probability per unit time is calculated in Appendix
C.. (Appendix D. is devoted to the approximations, identities and relations
which are used in the calculation of the cross section.)

\subsection{Cross section of electron assisted neutron exchange process}

The cross section $\sigma $ of the process can be obtained from the
transition probability per unit time $\left( \ref{Wfi22}\right) $ dividing
it by the flux $v_{e}/V$ \ of the incoming electron where $v_{e}$ is the
velocity of the electron. 
\begin{eqnarray}
\sigma &=&\int \frac{c}{v_{e}}\frac{\alpha _{f}^{2}\hbar
cZ^{2}\sum_{l_{2}=-m_{2}}^{l_{2}=m_{2}}\left\vert F_{2}\left( \mathbf{k}%
_{2}\right) \right\vert ^{2}}{\pi ^{3}v_{c}\left( \left\vert \mathbf{k}_{1}+%
\mathbf{k}_{2}\right\vert ^{2}+\lambda ^{2}\right) ^{2}\left( \Delta E_{\mu
i}\right) _{\mathbf{k}_{n}=\mathbf{k}_{2}}^{2}}  \label{sigma} \\
&&\times \frac{F_{e}(E_{ie})}{F_{e}(E_{f1})}\left\langle \left\vert
F_{1}\left( \mathbf{k}_{2}\right) \right\vert ^{2}\right\rangle
A_{2}^{2}r_{A_{2}}\delta (E_{f}-\Delta )d^{3}k_{1}d^{3}k_{2},  \notag
\end{eqnarray}%
where $v_{c}$ is the volume of elementary cell in the solid, $r_{A_{2}}$ is
the relative natural abundance of atoms $_{Z}^{A_{2}}X$, 
\begin{equation}
F_{1}\left( \mathbf{k}_{2}\right) =\int \Phi _{i1}\left( \mathbf{r}%
_{n1}\right) e^{-i\mathbf{k}_{2}\frac{A_{1}}{A_{1}-1}\cdot \mathbf{r}%
_{n1}}d^{3}r_{n1},  \label{F1kalk2}
\end{equation}%
\begin{equation}
\left\langle \left\vert F_{1}\left( \mathbf{k}_{2}\right) \right\vert
^{2}\right\rangle =\frac{1}{2l_{1}+1}\sum_{l_{1}=-m_{1}}^{l_{1}=m_{1}}\left%
\vert F_{1}\left( \mathbf{k}_{2}\right) \right\vert ^{2}  \label{F1av}
\end{equation}%
and%
\begin{eqnarray}
F_{2}\left( \mathbf{k}_{2}\right) &=&\int \Phi _{f2}^{\ast }\left( \mathbf{r}%
_{n2}\right) e^{i\mathbf{k}_{2}\cdot \mathbf{r}_{n2}}\times  \label{F2k2} \\
&&\times \left( -f\frac{\exp (-s\frac{A_{2}+1}{A_{2}}r_{n2}}{\frac{A_{2}+1}{%
A_{2}}r_{n2}}\right) d^{3}r_{n2}.  \notag
\end{eqnarray}%
Here $\Phi _{i1}$ and $\Phi _{f2}$ are the initial and final bound neutron
states (for the definition of $l_{1}$ and $l_{2}$ see below). The cross
section calculation result that the $k_{2}\simeq k_{0}=\sqrt{2\mu
_{12}\Delta }/\hbar $ substitution may be used (see in Appendix D.) in
calculating $F_{1}$ and $F_{2}$ in $\sigma $, where $\mu _{12}=m_{0}$ $\left[
\left( A_{1}-1\right) \left( A_{2}+1\right) \right] /\left(
A_{1}+A_{2}\right) $.

When evaluating $\left( \ref{sigma}\right) $ first the Weisskopf
approximation is applied, i.e. for the initial and final bound neutron
states we take $\Phi _{W}\left( \mathbf{r}_{nj}\right) =\phi _{jW}\left(
r_{nj}\right) Y_{l_{j}m_{j}}\left( \Omega _{j}\right) ,$ \ $j=1,2$ where $%
Y_{l_{j}m_{j}}\left( \Omega _{j}\right) $ is a spherical harmonics and $\phi
_{jW}\left( r_{nj}\right) =\sqrt{3/R_{j}^{3}},$ \ $j=1,2$ if $\left\vert 
\mathbf{r}_{nj}\right\vert \leq R_{j}$ and $\phi _{jW}\left( r_{nj}\right)
=0 $ for $\left\vert \mathbf{r}_{nj}\right\vert >R_{j}$, where $%
R_{j}=r_{0}A_{j}^{1/3}$is the radius of a nucleus of nucleon number $A_{j}$
with $r_{0}=1.2\times 10^{-13}$ $cm$. We apply the $A_{1}\simeq A_{2}\simeq
A_{1}-1\simeq A_{2}+1=A$ approximation further on. Calculating $F_{1}\left( 
\mathbf{k}_{0}\right) $ and $F_{2}\left( \mathbf{k}_{0}\right) $ the long
wavelength approximations (LWA) ($\exp \left( -i\mathbf{k}_{0}\cdot \mathbf{r%
}_{n1}\right) =1$ and $\exp \left( i\mathbf{k}_{0}\cdot \mathbf{r}%
_{n2}\right) =1$) are also used with $s=1/r_{0}$ that result approximately 
\begin{equation}
\left\langle \left\vert F_{1}\left( \mathbf{k}_{0}\right) \right\vert
^{2}\right\rangle \sum_{l_{2}=-m_{2}}^{l_{2}=m_{2}}\left\vert F_{2}\left( 
\mathbf{k}_{0}\right) \right\vert ^{2}=16\pi ^{2}r_{0}^{4}f^{2}\left(
2l_{2}+1\right) .  \label{F1K2av}
\end{equation}%
Using the results of Appendix D., the $E_{f1}=\Delta /2$ relation and if $%
E_{e}<0.1$ $MeV$ (i.e. if $F_{e}(E_{ie})=\left\vert 2\pi \eta _{e}\left(
E_{ie}\right) \right\vert =2\pi Z\alpha _{f}\sqrt{m_{e}c^{2}/2E_{ie}}$) then
the cross section in the Weisskopf-LWA approximation reads as 
\begin{equation}
\sigma _{W}=\frac{C_{W0}\left( 2l_{2}+1\right) }{\left[ 1+\frac{2\left(
\Delta _{n}-\Delta _{-}\right) }{A\Delta }\right] ^{2}}\frac{r_{A_{2}}}{%
F_{e}(\Delta /2)}\frac{A^{3/2}Z^{2}}{\Delta ^{3/2}E_{ie}}  \label{sigma2}
\end{equation}%
with $C_{W0}=2^{9}\pi ^{3}\alpha _{f}^{3}\left( 0.08\right)
^{2}a_{B}r_{0}\left( \frac{r_{0}}{d}\right) ^{3}\left( m_{0}c^{2}\right)
^{3/2}m_{e}c^{2}$. Here $a_{B}$ is the Bohr-radius, the relation $c/v_{e}=%
\sqrt{m_{e}c^{2}/\left( 2E_{ie}\right) }$ with $E_{ie}$ the kinetic energy
of the ingoing electrons is also applied and $d=3.52\times 10^{-8}$ $cm$ ($%
Ni $ lattice) and $d=3.89\times 10^{-8}$ $cm$ ($Pd$ lattice). $F_{e}(\Delta
/2)$ is determined by $\left( \ref{FeE}\right) $ and $\left( \ref{eatae}%
\right) $. The subscript $W$ refers to the Weisskopf-LWA approximation and
in $\left( \ref{sigma2}\right) $ the quantities $\Delta $ and $E_{ie}$ have
to be substituted in $MeV$ units. $C_{W0}\left( Ni\right) =1.4\times 10^{-14}
$ $MeV^{5/2}b$ and $C_{W0}\left( Pd\right) =1.1\times 10^{-14}$ $MeV^{5/2}b$.

We have calculated $\sum_{l_{2}=-m_{2}}^{l_{2}=m_{2}}\left\vert F_{2}\left( 
\mathbf{k}_{0}\right) \right\vert ^{2}$, $\left\langle \left\vert
F_{1}\left( \mathbf{k}_{0}\right) \right\vert ^{2}\right\rangle $ and the
cross section in the single particle shell model with isotropic harmonic
oscillator potential and without the long wavelength approximation (see
Appendix E.). We introduce the ratio 
\begin{equation}
\eta =\frac{\left\langle \left\vert F_{1}\left( \mathbf{k}_{0}\right)
\right\vert ^{2}\right\rangle
_{Sh}\sum_{l_{2}=-m_{2}}^{l_{2}=m_{2}}\left\vert F_{2}\left( \mathbf{k}%
_{0}\right) \right\vert _{Sh}^{2}}{\left\langle \left\vert F_{1}\left( 
\mathbf{k}_{0}\right) \right\vert ^{2}\right\rangle
_{W}\sum_{l_{2}=-m_{2}}^{l_{2}=m_{2}}\left\vert F_{2}\left( \mathbf{k}%
_{0}\right) \right\vert _{W}^{2}}.  \label{etha}
\end{equation}%
(The subscript $Sh$ refers to the shell model.) With the aid of $\eta \equiv
\eta _{l_{1},n_{1},l_{2},n_{2}}\left( A_{1},A_{2}\right) $ given by $\left( %
\ref{etha2}\right) $ (see Appendix E.) the cross section $\sigma _{Sh}$
calculated in the shell model can be written as 
\begin{equation}
\sigma _{Sh}=\eta _{l_{1},n_{1},l_{2},n_{2}}\left( A_{1},A_{2}\right) \sigma
_{W}.  \label{sigmaSh}
\end{equation}

\subsection{Yield of events of electron assisted neutron exchange process}

The yield $dN/dt$ of events of electron assisted neutron exchange process $%
A_{1},A_{2}\rightarrow A_{1}-1,A_{2}+1$ can be written as 
\begin{equation}
\frac{dN}{dt}=N_{t}N_{ni}\sigma \Phi ,  \label{rate1}
\end{equation}%
where $\sigma =\left\{ \sigma _{W}\text{ or }\sigma _{Sh}\right\} $, $\Phi $
is the flux of electrons, $N_{t}$ is the number of target particles, i.e.
the number $N_{A_{1}}$ of irradiated atoms of mass number $A_{1}$ in the
metal. The contribution of $N_{ni}$ neutrons in each nucleus $_{Z}^{A_{1}}X$
is also taken into account. $N_{ni}$ is the number of neutrons in the
uppermost energy level of the initial nucleus $_{Z}^{A_{1}}X$. If $F$ and $D$
are the irradiated surface and the width of the sample, respectively, then
the number of elementary cells $N_{c}$ in the sample is $%
N_{c}=FD/v_{c}=4FD/d^{3}$ in the case of $Ni$ and $Pd$, and the number of
atoms in the elementary cell is $2r_{A_{1}}$ with $r_{A_{1}}$ the relative
natural abundance of atoms $_{Z}^{A_{1}}X$ thus the number $N_{t}$ of target
atoms of mass number $A_{1}$ in the process is 
\begin{equation}
N_{t}=\frac{8}{d^{3}}r_{A_{1}}FD.  \label{NA1}
\end{equation}

The wave numbers and energies of the two outgoing heavy particles are
approximately $\mathbf{k}_{1}=-\mathbf{k}_{2}$,%
\begin{equation}
E_{1}=\frac{A_{2}+1}{A_{1}+A_{2}}\Delta \text{ and }E_{2}=\frac{A_{1}-1}{%
A_{1}+A_{2}}\Delta .  \label{E1}
\end{equation}

\subsection{Numerical data of electron assisted neutron exchange processes
in $Ni$ and $Pd$}

\begin{table}[tbp]
\tabskip=8pt 
\centerline {\vbox{\halign{\strut $#$\hfil&\hfil$#$\hfil&\hfil$#$
\hfil&\hfil$#$\hfil&\hfil$#$\hfil&\hfil$#$\hfil&\hfil$#$\cr
\noalign{\hrule\vskip2pt\hrule\vskip2pt}
A &58 &60 &61 &62 &64 \cr
\Delta_{-} &-4.147 &-3.317 &0.251 &-2.526 &-1.587 \cr
\Delta_{+} &0.928 &-0.251 &2.526 &-1.234 &-1.973 \cr
r_{A} &0.68077 &0.26223 &0.0114 &0.03634 &0.00926 \cr
\noalign{\vskip2pt\hrule\vskip2pt\hrule}}}}
\caption{Numerical data of the $\text{ }e+\text{ }_{28}^{A_{1}}Ni+\text{ }%
_{28}^{A_{2}}Ni\rightarrow e^{\prime }+\text{ }_{28}^{A_{1}-1}Ni+\text{ }%
_{28}^{A_{2}+1}Ni+\Delta $ reaction. The reaction is energetically allowed
if $\Delta =\Delta _{-}(A_{1})+\Delta _{+}(A_{2})>0$ holds. $A$ is the mass
number, $r_{A}$ is the relative natural abundance, $\Delta _{-}(A)=\Delta
_{A}-\Delta _{A-1}$ and $\Delta _{+}(A)=\Delta _{A}-\Delta _{A+1}$ are given
in $MeV$ units.}
\label{Table1}
\end{table}

\begin{table}[tbp]
\tabskip=8pt 
\centerline {\vbox{\halign{\strut $#$\hfil&\hfil$#$\hfil&\hfil$#$
\hfil&\hfil$#$\hfil&\hfil$#$\hfil&\hfil$#$\hfil&\hfil$#$\cr
\noalign{\hrule\vskip2pt\hrule\vskip2pt}
A &102 &104 &105 &106 &108 &110 \cr
\Delta_{-} &-2.497 &-1.912 &0.978 &-1.491 &-1.149 &-0.747 \cr
\Delta_{+} &-0.446 &-0.978 &1.491 &-1.533 &-1.918 &-2.320 \cr
r_{A} &0.0102 &0.1114 &0.2233 &0.2733 &0.2646 &0.1172 \cr
\noalign{\vskip2pt\hrule\vskip2pt\hrule}}}}
\caption{Numerical data of the $\text{ }e+\
_{46}^{A_{1}}Pd+_{46}^{A_{2}}Pd\rightarrow e^{\prime }+\text{ }%
_{46}^{A_{1}-1}Pd+_{46}^{A_{2}+1}Pd+\Delta$ reaction. The reaction is
energetically allowed if $\Delta=\Delta_{-}(A_{1})+\Delta_{+}(A_{2})>0$
holds. $A$ is the mass number, $r_{A}$ is the relative natural abundance, $%
\Delta_{-}(A)=\Delta_{A}-\Delta_{A-1} $ and $\Delta_{+}(A)=\Delta_{A}-%
\Delta_{A+1} $ are given in $MeV$ units.}
\label{Table2}
\end{table}

\begin{table}[tbp]
\tabskip=8pt 
\centerline {\vbox{\halign{\strut $#$\hfil &\hfil$#$\hfil&\hfil$#$
\hfil&\hfil$#$\hfil\cr
\noalign{\hrule\vskip2pt\hrule\vskip2pt}
 A_{1}\rightarrow A_{1}-1&A_{2}\rightarrow A_{2}+1&\Delta($MeV$)&\eta \cr
\noalign{\vskip2pt\hrule\vskip2pt}
61 \rightarrow 60 &58 \rightarrow 59 & 1.179 &7.02\times10^{-3}\cr
61 \rightarrow 60 &61 \rightarrow 62 & 2.777 &2.42\times10^{-8}\cr
64 \rightarrow 63 &61 \rightarrow 62 & 0.939 &2.08\times10^{-4}\cr
\noalign{\vskip2pt\hrule\vskip2pt\hrule}}}}
\caption{The values of the quantities $\protect\eta$ and $\Delta =\Delta
_{-}(A_{1})+\Delta _{+}(A_{2})>0$, the later in $MeV $ units, of the $\text{ 
}e+\text{ }_{28}^{A_{1}}Ni+\text{ }_{28}^{A_{2}}Ni\rightarrow e^{\prime }+%
\text{ }_{28}^{A_{1}-1}Ni+\text{ }_{28}^{A_{2}+1}Ni+\Delta $ reaction. The $%
\Delta _{-}(A_{1})$ and $\Delta _{+}(A_{2})$ values can be found in Table I.
For the definition of $\protect\eta $ see $\left( \protect\ref{etha}\right)$
and $\left( \protect\ref{etha2}\right)$.}
\label{Table3}
\end{table}

\begin{table}[tbp]
\tabskip=8pt 
\centerline {\vbox{\halign{\strut $#$\hfil &\hfil$#$\hfil&\hfil$#$
\hfil&\hfil$#$\hfil\cr
\noalign{\hrule\vskip2pt\hrule\vskip2pt}
 A_{1}\rightarrow A_{1}-1&A_{2}\rightarrow A_{2}+1&\Delta($MeV$)&\eta \cr
\noalign{\vskip2pt\hrule\vskip2pt}
105 \rightarrow 104 &102 \rightarrow 103 & 0.532 & 1.84\times10^{-4}\cr
105 \rightarrow 104 &105 \rightarrow 106 & 2.469 & 8.88\times10^{-11}\cr
108 \rightarrow 107 &105 \rightarrow 106 & 0.342 & 2.82\times10^{-3}\cr
\noalign{\vskip2pt\hrule\vskip2pt\hrule}}}}
\caption{The values of the quantities $\protect\eta$ and $\Delta =\Delta
_{-}(A_{1})+\Delta _{+}(A_{2})>0$, the later in $MeV $ units, of the $\text{ 
}e+\ _{46}^{A_{1}}Pd+_{46}^{A_{2}}Pd\rightarrow e^{\prime }+\text{ }%
_{46}^{A_{1}-1}Pd+_{46}^{A_{2}+1}Pd+\Delta $ reaction. The $\Delta
_{-}(A_{1})$ and $\Delta _{+}(A_{2})$ values can be found in Table II. For
the definition of $\protect\eta$ see $\left( \protect\ref{etha}\right)$ and $%
\left( \protect\ref{etha2}\right)$.}
\label{Table4}
\end{table}

As a first example we take $Ni$ as target material. In this case the
possible processes are%
\begin{equation}
\text{ }e+\text{ }_{28}^{A_{1}}Ni+\text{ }_{28}^{A_{2}}Ni\rightarrow
e^{\prime }+\text{ }_{28}^{A_{1}-1}Ni+\text{ }_{28}^{A_{2}+1}Ni+\Delta .
\label{NiAp}
\end{equation}%
Tables I. and III. contain the relevant data for reaction $\left( \ref{NiAp}%
\right) $. Describing neutrons in the uppermost energy level of $_{28}^{A}Ni$
isotopes we used $1p$ shell model states in the cases of $A=58-60$ and $0f$
shell model states in the cases of $A=61-64$.

Another interesting target material is $Pd$ in which the electron assisted
neutron exchange processes are the 
\begin{equation}
\text{ }e+\ _{46}^{A_{1}}Pd+\text{ }_{46}^{A_{2}}Pd\rightarrow e^{\prime }+%
\text{ }_{46}^{A_{1}-1}Pd+\text{ }_{46}^{A_{2}+1}Pd+\Delta  \label{PdAp}
\end{equation}%
reactions. The relevant data can be found in Tables II. and IV.. Describing
neutrons in the uppermost energy level of $_{46}^{A}Pd$ isotopes we used $0g$
shell model states in the cases of $A=102-104$ and $1d$ shell model states
in the cases of $A=105-108$. The nuclear data to the Tables are taken from 
\cite{Shir}. One can see from Tables III. and IV. that in both cases three
possible pairs of isotopes exist which are energetically allowed (for which $%
\Delta >0$) and their rates differ in the factor $\left( 2l_{2}+1\right)
N_{ni}\eta _{l_{1},n_{1},l_{2},n_{2}}\left( A_{1},A_{2}\right)
r_{A_{1}}r_{A_{2}}\Delta ^{-3/2}$ only. The $\eta \equiv \eta
_{l_{1},n_{1},l_{2},n_{2}}\left( A_{1},A_{2}\right) $ values of $Ni$ and $Pd$
can also be found in Tables III. and IV., respectively. The results of
numerical investigation of $\left( 2l_{2}+1\right) N_{ni}\eta
_{l_{1},n_{1},l_{2},n_{2}}\left( A_{1},A_{2}\right) r_{A_{1}}r_{A_{2}}\Delta
^{-3/2}$ shows that the $61\rightarrow 60,58\rightarrow 59$ and the $%
108\rightarrow 107,105\rightarrow 106$ reactions are the dominant among the
processes in $Ni$ and $Pd$, respectively.

In the case of $Ni$ it is found that the 
\begin{equation}
\text{ }e+\text{ }_{28}^{61}Ni+\text{ }_{28}^{58}Ni\rightarrow e^{\prime }+%
\text{ }_{28}^{60}Ni+\text{ }_{28}^{59}Ni+1.179\text{ }MeV  \label{Nilead}
\end{equation}%
process of $\sigma _{Sh}=0.088/E_{ie}$ $\mu b$ with $E_{ie}$ in $MeV$ is
leading. In this case the $_{28}^{60}Ni$ and the $_{28}^{59}Ni$ isotopes
take away $0.585$ $MeV$ and $0.594$ $MeV$, respectively. In the case of $Pd$
the 
\begin{equation}
e+\text{ }_{46}^{108}Pd+\text{ }_{46}^{105}Pd\rightarrow e^{\prime }+\text{ }%
_{46}^{107}Pd+\text{ }_{46}^{106}Pd+0.342\text{ }MeV  \label{Pdlead}
\end{equation}%
reaction of $\sigma _{Sh}=0.26/E_{ie}$ $\mu b$ with $E_{ie}$ in $MeV$ is
found to be the leading one. In this case the $_{46}^{107}Pd$ and the $%
_{46}^{106}Pd$ isotopes take away $0.170$ $MeV$ and $0.172$ $MeV$,
respectively.

\section{Other results - Other charged particle assisted reactions}

The transition probability per unit time and the cross section of the
processes, which will be discussed below, may be determined in similar
manner as was done above in the case of electron assisted neutron exchange
process. The main difference is that in matrix elements $V_{\mu i}^{Cb}$ and 
$V_{f\mu }^{St}$ different Coulomb factors appear according to the particles
which take part in the reaction.

\subsection{Electron assisted heavy charged particle exchange process}

There is an other possibility in the family of electron assisted exchange
processes, when a charged heavy particle (such as $p$, $d$, $t$, $_{2}^{3}He$
and $_{2}^{4}He$) is exchanged. The process is called electron assisted
heavy charged particle exchange process and it can be visualized with the
aid of Fig.1 too. Denoting the intermediate particle (particle $3$ in Fig.
1) by $_{z_{3}}^{A_{3}}w$, which is exchanged, the general electron assisted
heavy charged particle exchange processes reads as%
\begin{equation}
e+\text{ }_{Z_{1}}^{A_{1}}X+\text{ }_{Z_{2}}^{A_{2}}Y\rightarrow e^{\prime }+%
\text{ }_{Z_{1}-z_{3}}^{A_{1}-A_{3}}X^{\ast }+\text{ }%
_{Z_{2}+z_{3}}^{A_{2}+A_{3}}Y^{\ast }+\Delta .  \label{hpexchange}
\end{equation}%
Here $e$ and $e^{\prime }$ denote electron and $\Delta $ is the energy of
the reaction, i.e. the difference between the rest energies of initial $%
\left( _{Z_{1}}^{A_{1}}X+_{Z_{2}}^{A_{2}}Y\right) $ and final $\left(
_{Z_{1}-z_{3}}^{A_{1}-A_{3}}X^{\ast }+\text{ }%
_{Z_{2}+z_{3}}^{A_{2}+A_{3}}Y^{\ast }\right) $ states. $\Delta =\Delta
_{-}+\Delta _{+},$ with $\Delta _{-}=\Delta _{Z_{1}}^{A_{1}}-\Delta
_{Z_{1}-z_{3}}^{A_{1}-A_{3}}$ and $\Delta _{+}=\Delta
_{Z_{2}}^{A_{2}}-\Delta _{Z_{2}+z_{3}}^{A_{2}+A_{3}}$. $\Delta
_{Z_{1}}^{A_{1}}$, $\Delta _{Z_{1}-z_{3}}^{A_{1}-A_{3}}$ , $\Delta
_{Z_{2}}^{A_{2}}$, $\Delta _{Z_{2}+z_{3}}^{A_{2}+A_{3}}$ are the energy
excesses of neutral atoms of mass number-charge number pairs $A_{1}$, $Z_{1}$%
; $A_{1}-A_{3}$, $Z_{1}-z_{3}$; $A_{2}$, $Z_{2}$; $A_{2}+A_{3}$, $%
Z_{2}+z_{3} $, respectively \cite{Shir}.

In $\left( \ref{hpexchange}\right) $ the electron (particle $1$) Coulomb
interacts with the nucleus $_{Z_{1}}^{A_{1}}X$ (particle $2$). A scattered
electron (particle $1^{\prime }$), the intermediate particle $%
_{z_{3}}^{A_{3}}w$ (particle $3$) and the nucleus $%
_{Z_{1}-z_{3}}^{A_{1}-A_{3}}X^{\ast }$ (particle $2^{\prime }$) are created
due to this interaction. The intermediate particle $_{z_{3}}^{A_{3}}w$
(particle $3$) is captured due to the strong interaction by the nucleus $%
_{Z_{2}}^{A_{2}}Y$ (particle $4$) forming the nucleus $%
_{Z_{2}+z_{3}}^{A_{2}+A_{3}}Y^{\ast }$ (particle $5$) in this manner. So in $%
\left( \ref{hpexchange}\right) $ the nucleus $_{Z_{1}}^{A_{1}}X$ (particle $%
2 $) looses a particle $_{z_{3}}^{A_{3}}w$ which is taken up by the nucleus $%
_{Z_{2}}^{A_{2}}Y$ (particle $4$). The process is energetically forbidden if 
$\Delta <0$. Since particles $2^{\prime }$, $3$ and $4$ all have positive
charge, furthermore they all are heavy, the two Coulomb factors, which
appear in the cross section, are $F_{2^{\prime }3}$ and $F_{34}$. Therefore
the cross section of process $\left( \ref{hpexchange}\right) $ is expected
to be much smaller than the cross section of process $\left( \ref{exchange}%
\right) $. However process $\left( \ref{hpexchange}\right) $ may play an
essential role in explaining nuclear transmutations stated \cite{Storms2}
(see below). Since Coulomb factors $F_{2^{\prime }3}$ and $F_{34}$ determine
the order of magnitude of the cross section of the process (the cross
section of the process is proportional to $F_{2^{\prime }3}F_{34}$) we treat
them in more detail in Appendix F.

\subsection{Electron assisted nuclear capture process}

\begin{figure}[tbp]
\resizebox{6.0cm}{!}{\includegraphics*{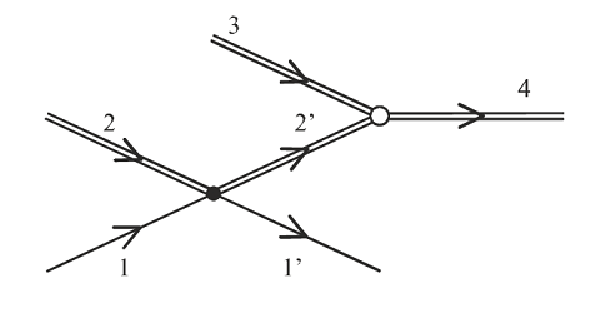}}
\caption{The graph of electron assisted nuclear capture reactions. The
simple lines represent free (initial (1) and final (1')) electrons. The
doubled lines represent free, heavy, charged initial (2) particles (such as
p, d), their intermediate state (2'), target nuclei (3) and reaction product
(4). The filled dot denotes Coulomb-interaction and the open circle denotes
nuclear (strong) interaction.}
\label{figure2}
\end{figure}

Now the electron assisted nuclear caption process (see Fig. 2) is
considered, in which an electron-nucleus Coulomb scattering is followed by a
capture process governed by strong interaction \cite{kk1}. When describing
the effect of the Coulomb interaction between the nucleus of charge number $%
Z $ and a slow electron one can also use the Coulomb factor $F_{e}(E)$ $(\ref%
{FeE})$ of the electron defined above.

As an example we consider the electron assisted $d+d\rightarrow $ $%
_{2}^{4}He $ process with slow deuterons. In this case, one of the slow
deuterons (as particle $2$) can enter into Coulomb interaction with a
quasi-free, slow electron (as particle $1$) of the solid (see Fig. 2). In
Coulomb scattering of free deuterons and electrons the wave number vector
(momentum) is preserved since their relative motion may be described by a
plane wave which is multiplied by the corresponding Coulomb factor. In this
second order process the Coulomb interaction is followed by strong
interaction, which induces a nuclear capture process. The energy $\Delta $
of the nuclear reaction is divided between the electron and the heavy
nuclear product. Since $m_{N}\gg m_{e}$ ($m_{N}$ is the rest mass of the
nuclear product), the electron will take almost all the total nuclear
reaction energy $\Delta $ away (there is no gamma emission) and the
magnitude $k_{1^{\prime }}$ of its wave number vector $\mathbf{k}_{1^{\prime
}}$ reads $k_{1^{\prime }}=\sqrt{\Delta ^{2}+2m_{e}c^{2}\Delta }/\left(
\hbar c\right) \simeq \Delta /\left( \hbar c\right) $ $\left( \text{if }%
\Delta \gg m_{e}c^{2}\right) $. If initially the electron and the deuteron
move slowly and the magnitudes of their wave number vectors are much smaller
than $\Delta /\left( \hbar c\right) $, then the initial wave number vectors
can be neglected in the wave number vector (momentum) conservation and
consequently, in the intermediate state (in state $2^{\prime }$) the
deuteron gets a wave number vector $\mathbf{k}_{2^{\prime }}=-\mathbf{k}%
_{1^{\prime }}$. If $\Delta =23.84$ $MeV$, which is the reaction energy of
the $d+d\rightarrow $ $_{2}^{4}He$ reaction, then the deuteron $2^{\prime }$
will have $k_{2^{\prime }}=\Delta /\left( \hbar c\right) $ and its
corresponding (virtual) kinetic energy $E_{2^{\prime }}=\Delta ^{2}/\left(
4m_{0}c^{2}\right) =76.5$ $keV$ in the $CM$ coordinate system. At this
energy the Coulomb factor value between particles $2^{\prime }$and $3$ reads
as $F_{2^{\prime }3}=0.103$. It must be compared to the extremely small
Coulomb factor value, e.g. in the case of energy $E=1$ $eV$ to $F_{23}\left(
1\text{ }eV\right) =1.1\times 10^{-427}$, that is characteristic of the
usual, first order process. If one compares again the cross sections of
second order and first order (electron assisted and usual) processes then
their ratio is approximately proportional to $F_{e}F_{2^{\prime
}3}/F_{23}(E) $ that becomes extremely large with decreasing $E$ too. (The
model and the details of calculation, the results and their discussion can
be found in \cite{kk1}.) The cross section of the electron assisted neutron
exchange process is expected to be larger than the cross section of electron
assisted nuclear capture process because of the appearance of the Coulomb
factor in it.

\subsection{Heavy particle assisted nuclear processes}

\begin{figure}[tbp]
\resizebox{6.0cm}{!}{\includegraphics*{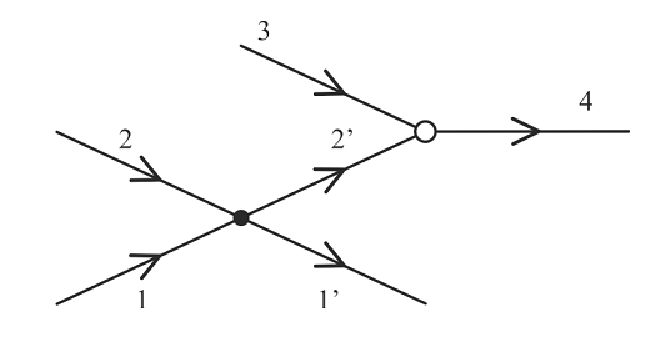}}
\caption{The graph of heavy particle assisted nuclear capture reactions. The
lines 1, 1' represent free (initial (1) and final (1')) heavy particle which
assists the reaction. The other lines represent heavy, charged initial (2)
particles, their intermediate state (2'), target nuclei (3) and reaction
product (4). The filled dot denotes Coulomb-interaction and the open circle
denotes nuclear (strong) interaction.}
\label{figure3}
\end{figure}

\begin{figure}[tbp]
\resizebox{7.0cm}{!}{\includegraphics*{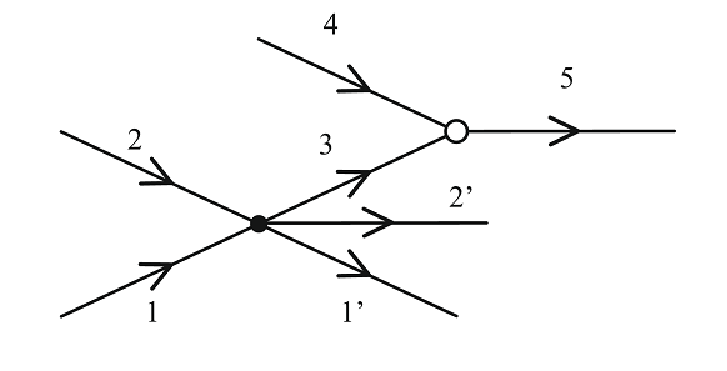}}
\caption{The graph of heavy particle assisted heavy charged particle (such
as $p$, $d$, $t$, $_{2}^{3}He$ and $_{2}^{4}He$) exchange reaction. The
lines 1, 1' represent free (initial (1) and final (1')) heavy particle which
assists the reaction. The other lines represent heavy, charged initial
nuclei (2), their final state (2', which is a nucleus lost particle 3), the
transferred particle (3), target nuclei (4) and reaction product (5). The
filled dot denotes Coulomb-interaction and the open circle denotes nuclear
(strong) interaction.}
\label{figure4}
\end{figure}

In electron assisted nuclear reactions heavy, charged particles of energy of
a few $MeV$ may be created. In the decelerating process of reaction products
of the electron assisted processes the energy of these heavy particles may
become intermediately low (of about $0.01$ $\left[ MeV\right] $) so their
Coulomb factor, if the particles are light, may be intermediately small so
their assistance in nuclear processes have to be also considered among the
accountable nuclear processes. The corresponding graphs can be seen in Fig.
3 and Fig. 4. Fig. 3 depicts a heavy, charged particle assisted nuclear
capture process and Fig. 4 represents heavy, charged particle assisted heavy
charged particle (such as $p$, $d$, $t$, $_{2}^{3}He$ and $_{2}^{4}He$)
exchange reaction. Now all particles are heavy. According to the applied
notation, particles $2^{\prime }$, $3$ (in Fig. 3) and particles $3$, $4$
(in Fig. 4) take part in a nuclear process and particle $1$ only assists it.
The different processes will be distinguished by the type of the assisting
particle and also by the type of the nuclear process. In our model charged,
heavy particles, such as protons $\left( p\right) $, deuterons $\left(
d\right) $ may be particle $1$, which are supposed to move freely in a solid
(e.g. in a metal). The other particles, that may take part in the processes
are: localized heavy, charged particles (bound, localized $p$, $d$ and other
nuclei) as the participants of Coulomb scattering (with particle $1$) and
localized heavy, charged particles (bound, localized $p$, $d$ and other
nuclei) as nuclear targets (as particle $3$ in Fig. 3 and particle $4$ in
Fig. 4). The problem, that there may be identical particles in the system
that are indistinguishable, is also disregarded here.

The calculation of the transition probability per unit time of the process
can be performed through similar steps to those applied for the calculation
of the rate of an electron assisted process. The main difference is that now
particle $1$ is heavy. In order to show the capability of the heavy particle
assisted nuclear processes, some cases of the proton assisted proton
captures 
\begin{equation}
p+\text{ }_{Z}^{A}X+p\rightarrow \text{ }_{Z+1}^{A+1}Y+p^{\prime }+\Delta
\end{equation}%
were investigated in Appendix III. (Ch. IX.) of \cite{kk1}.

\section{Discussion - Analysis of experimental observations}

\subsection{Fleischmann-Pons experiment}

In the experiment of \cite{FP1} $Pd$ was filled with deuterons during
electrolysis. The electrolyte had $LiOD$ content too. Two types of electron
assisted neutron exchange processes with $Pd$ nuclei are possible:%
\begin{equation}
e+d+_{46}^{A}Pd\rightarrow e^{\prime }+p+_{46}^{A+1}Pd+\Delta  \label{Pd1}
\end{equation}%
with $\Delta =\Delta _{-}(d)+\Delta _{+}(A)$ and 
\begin{equation}
e+_{3}^{7}Li+_{46}^{A}Pd\rightarrow e^{\prime
}+_{3}^{6}Li+_{46}^{A+1}Pd+\Delta  \label{Pd2}
\end{equation}%
with $\Delta =\Delta _{-}(Li)+\Delta _{+}(A)$ (the $\Delta _{+}(A)$ values
can be found in Table III). $\Delta _{-}(d)=\Delta _{d}-\Delta _{p}=5.847$ $%
MeV$ and $\Delta _{-}(Li)=\Delta (_{3}^{7}Li)-\Delta (_{3}^{6}Li)=0.821$ $%
MeV $ are the energies of neutron loss of $d$ and $_{3}^{7}Li$, where $%
\Delta _{d}$, $\Delta _{p}$, $\Delta (_{3}^{7}Li)$ and $\Delta (_{3}^{6}Li)$
are the mass excesses of deuteron, proton, $_{3}^{7}Li$ and $_{3}^{6}Li$,
respectively. In reactions $\left( \ref{Pd1}\right) $ and $\left( \ref{Pd2}%
\right) $ electrons of the metal are particle $1$, $d$ and $_{3}^{7}Li$ are
particle $2$ and $_{46}^{A}Pd$ appears as particle $4$ (see Fig. 1).
Reaction $\left( \ref{Pd1}\right) $ is energetically allowed for all the
natural isotopes of $Pd$ since $\Delta =\Delta _{-}(d)+\Delta _{+}(A)>0$ for
each $A$ (see the $\Delta _{+}(A)$ values of Table II). In the case of
reaction $\left( \ref{Pd2}\right) $ the $\Delta =\Delta _{-}(Li)+\Delta
_{+}(A)>0$ condition holds at $A=102$ and $A=105$ resulting $\Delta =0.375$ $%
MeV$ \ and $\Delta =2.312$ $MeV$, respectively.

However, at the $Pd$ surface other types of electron assisted neutron
exchange processes with $d$ and $Li$ nuclei of the electrolyte and $d$
solved in $Pd$ are possible:%
\begin{equation}
e+\text{ }d+d\rightarrow e^{\prime }+p+\text{ }t+\Delta ,  \label{ddtp}
\end{equation}%
\begin{equation}
e+\text{ }d+d\rightarrow e^{\prime }+n+\text{ }_{2}^{3}He+\Delta ,
\label{ddnHe}
\end{equation}%
\begin{equation}
e+d+\text{ }_{3}^{6}Li\rightarrow e^{\prime }+p+\text{ }_{3}^{7}Li+\Delta ,
\label{Li1}
\end{equation}%
\begin{equation}
e+d+\text{ }_{3}^{6}Li\rightarrow e^{\prime }+2_{2}^{4}He+\Delta ,
\label{Li2}
\end{equation}%
\begin{equation}
e+d+\text{ }_{3}^{7}Li\rightarrow e^{\prime }+2_{2}^{4}He+n+\Delta
\label{Li3}
\end{equation}%
and%
\begin{equation}
e+d+\text{ }_{3}^{7}Li\rightarrow e^{\prime }+\text{ }_{4}^{8}Be+n+\Delta ,
\label{Li4}
\end{equation}%
which is promptly followed by the decay $_{4}^{8}Be\rightarrow 2_{2}^{4}He$ (%
$\Gamma _{\alpha }=6.8$ $eV$). In these reactions electrons of the metal are
particle $1$ and $d$ is particle $2$.

In reaction $\left( \ref{Pd1}\right) $ protons of energy up to $7.269$ $MeV$
and in reaction $\left( \ref{Pd2}\right) $ $_{3}^{6}Li$ particles of maximum
energy $2.189$ $MeV$ are created which may enter into usual nuclear
reactions with the nuclei of deuteron loaded $Pd$ and electrolyte which are
(without completeness): the usual $pd\rightarrow $ $_{2}^{3}He+\gamma $
reaction,%
\begin{equation}
p+_{3}^{7}Li\rightarrow 2_{2}^{4}He+Q\text{ with }Q=\Delta +E_{kin}(p),
\label{pLi}
\end{equation}%
\begin{equation}
_{3}^{6}Li+d\rightarrow 2_{2}^{4}He+Q\text{ with }Q=\Delta +E_{kin}(Li),
\label{dLi1}
\end{equation}%
\begin{equation}
_{3}^{6}Li+d\rightarrow p+_{3}^{7}Li+Q\text{ with }Q=\Delta +E_{kin}(Li).
\label{dLi2}
\end{equation}%
In $\left( \ref{pLi}\right) $ and $\left( \ref{dLi1}\right) $ the emitted $%
_{2}^{4}He$ has energy $E_{^{4}He}>8.674$ $MeV$ and $E_{^{4}He}>11.186$ $MeV$%
, and in $(\ref{dLi2})$ the created $p$ and $_{3}^{7}Li$ have energy $%
E_{p}>4.397$ $MeV$ and $E_{^{7}Li}>0.628$ $MeV$, respectively. It can be
seen that in $\left( \ref{pLi}\right) $ and $\left( \ref{dLi1}\right) $ $%
_{2}^{4}He$ is produced. The $_{3}^{7}Li$ particles may enter into reaction%
\begin{equation}
_{3}^{7}Li+d\rightarrow 2_{2}^{4}He+n+Q\text{ with }Q=\Delta +E_{kin}(Li)%
\text{ }  \label{dLi3}
\end{equation}%
which contributes to the $_{2}^{4}He$ production too. Here and above $%
E_{kin}(p)$ and $E_{kin}(Li)$ are the kinetic energies of the initial
protons, $_{3}^{6}Li$ and $_{3}^{7}Li$ isotopes.

From the above one can see that at least twelve types of reactions
(altogether $18$ reactions) exist which are capable of energy production and
in half of them energy production is accompanied with $_{2}^{4}He$
production. It is reasonable that reactions $\left( \ref{Pd1}\right) $ and $%
\left( \ref{Pd2}\right) $ have the highest rate in the above list of
reactions. In the majority of the above reactions charged particles, mostly
heavy charged particles are created with short range and so they loose their
energy in the matter of the experimental apparatus mainly in the electrode
(cathode) and the electrolyte, therefore their direct observation is
difficult. It is mainly heat production, which is a consequence of
deceleration in the matter of the apparatus, that can be experienced. The
third of the processes, mainly the secondary processes are the sources of
neutron emission. X- and $\gamma -$rays may be originated mainly from
bremsstrahlung. The above reasoning tallies with experimental observations.

In reactions $\left( \ref{Pd1}\right) -\left( \ref{dLi3}\right) $ heavy,
charged particles of kinetic energy lying in the $MeV$ range are created
which are able to assist nuclear reactions. One can obtain the possible
heavy charged particles assisted reactions if in reactions $\left( \ref{Pd1}%
\right) -\left( \ref{Li4}\right) $ the electron is replaced by heavy charged
particles ($p$, $t$, $_{2}^{3}He$, $_{2}^{4}He$, $_{3}^{6}Li$, $_{3}^{7}Li$, 
$_{4}^{8}Be$ and $_{46}^{A+1}Pd$ with $A=102,104-106,108,110$) which are
created in reactions $\left( \ref{Pd1}\right) -\left( \ref{dLi3}\right) $.
Since the number of possible heavy charged particles is $13$ and the number
of reactions which may be assisted by them is $8$, at least $104$ heavy
charged particle assisted reactions must be taken into account.
Consequently, it is a rather great theoretical challenge and task to
determine precisely the relative rates and their couplings of all the
accountable reactions, a work which is, nevertheless, necessary for the
accurate quantitative analysis of experiments.

The relative rates of coupled reactions of many types depend significantly
on the geometry, the kind of matter and other parameters of the experimental
apparatus and on some further variables, which may be attached to a concrete
experiment. This situation may be responsible for the diversity of the
results of experiments, which are thought to have been carried out with
seemingly in the same circumstances.

\subsection{Nuclear transmutation}

As to the phenomenon of nuclear transmutation \cite{Storms2} we demonstrate
its possibility only. First let us see the possibility of normal reactions.
For instance in a Fleischmann-type experiment $_{3}^{6}Li$ particles of
energy up to $2.189$ $MeV$ are created in reaction $\left( \ref{Pd2}\right) $
so the reaction%
\begin{equation}
_{3}^{6}Li+_{3}^{6}Li\rightarrow _{6}^{12}C+\gamma +Q\text{ }  \label{LiLI}
\end{equation}%
may have minor, but measurable probability. Here $Q=\Delta +E_{kin}(Li)$.

The Coulomb factor of reaction $\left( \ref{LiLI}\right) $ is $%
F_{Li,Li}=1.71\times 10^{-3}$ at $2.189$ $MeV$ kinetic energy of $_{3}^{6}Li$
particles. The magnitude of the Coulomb factor indicates that the rate of
reaction $\left( \ref{LiLI}\right) $ may be large enough to be able to
produce carbon traces in observable quantity.

Moreover, in reactions $\left( \ref{Pd1}\right) $ and $\left( \ref{Pd2}%
\right) $ free $_{46}^{A}Pd$ particles are created offering e.g. the
possibility of the 
\begin{equation}
e+_{46}^{A_{1}}Pd+_{46}^{A_{2}}Pd\rightarrow e^{\prime
}+_{44}^{A_{1}-3}Ru+_{48}^{A_{2}+3}Cd+\Delta  \label{He3exchange}
\end{equation}%
electron assisted $_{2}^{3}He$ exchange process. The electron and the other $%
Pd$ particle are in the solid. Analyzing mass excess data \cite{Shir} it was
found that e.g. the $e+_{46}^{103}Pd+_{46}^{111}Pd\rightarrow e^{\prime
}+_{44}^{100}Ru+_{48}^{114}Cd+\Delta $ $\ _{2}^{3}He$ exchange process has
reaction energy $\Delta =5.7305$ $MeV$. [$_{46}^{103}Pd$ and $_{46}^{111}Pd$
are produced in reaction $\left( \ref{Pd1}\right) $.] Calculating the $%
F_{2^{\prime }3}=F_{34}$ Coulomb factors taking $A=100$, $Z=46$, $A_{3}=3$,$%
\ z_{3}=2$ in $\left( \ref{F2'3F34}\right) $ one gets $F_{2^{\prime
}3}F_{34}=2.5\times 10^{-12}$ which seems to be large enough number to
produce $Cd$ and $Ru$ traces in an experiment lasting many days long.

The above reactions may offer\ starting point for the explanation of nuclear
transmutations.

\subsection{Rossi-type reactor (E-cat)}

Recently the Rossi-type reactor \cite{Rossi} (E-Cat) was experimentally
investigated in detail \cite{Levi}. The fuel contained mostly $Ni$ and also $%
Li$ in accountable measure, there was $0.011g$ $Li$ in $1$ $g$ fuel. The
isotope composition of the unused fuel was equal to the relative natural
abundances. But the isotope composition of the ash (the fuel after 32 day
run of the reactor) strongly changed. (The measured relative abundances of $%
Li$ and $Ni$ isotopes in fuel and ash can be seen in Table V. The natural
abundances are also given for comparison. The data are taken from Appendix
3. of \cite{Levi}.) One can see that the $_{28}^{62}Ni$ isotope is enriched
and the other $Ni$ isotopes are depleted. Furthermore, the relative $%
_{3}^{7}Li$ content decreased from $0.917$ to $0.079$ while the relative $%
_{3}^{6}Li$ content increased from $0.086$ to $0.921$.

\begin{table}[tbp]
\tabskip=8pt 
\centerline {\vbox{\halign{\strut $#$\hfil &\hfil$#$\hfil&\hfil$#$
\hfil&\hfil$#$\hfil\cr
\noalign{\hrule\vskip2pt\hrule\vskip2pt}
 Isotope&Fuel&Ash&Natural \cr
\noalign{\vskip2pt\hrule\vskip2pt}
_{3}^{6}Li  & 0.086 & 0.921 & 0.075\cr
_{3}^{7}Li  & 0.914 & 0.079 & 0.925\cr
_{28}^{58}Ni & 0.67 & 0.008 & 0.681\cr
_{28}^{60}Ni & 0.263 & 0.005 & 0.262\cr
_{28}^{61}Ni & 0.019 & 0.000 & 0.018\cr
_{28}^{62}Ni & 0.039 & 0.987 & 0.036\cr
_{28}^{64}Ni & 0.01 & 0 & 0.009\cr
\noalign{\vskip2pt\hrule\vskip2pt\hrule}}}}
\caption{Measured relative abundances of $Li$ and $Ni$ isotopes in fuel and
ash. The natural relative abundances are also given for comparison. The data
are taken from \protect\cite{Levi}.}
\label{Table5}
\end{table}

The reactor worked for about ten days at temperature $T_{1}=1533$ $K$ and
the remaining time at temperature $T_{2}=1673$ $K$. At these temperatures a
free electron gas may be created from the $Ni$ powder of the fuel due to the
termionic emission process. The emitted flux of electrons can be determined
from the current density of electrons according to the Richardson's law
using the work function $U=5.24$ $eV$ of $Ni$. The obtained termionic
electron fluxes are $\Phi _{1}=$ $7.5\times 10^{9}$ $cm^{-2}s^{-1}$and $\Phi
_{2}=$ $2.4\times 10^{11}$ $cm^{-2}s^{-1}$ at $T_{1}$ and $T_{2}$,
respectively. Regarding the large surface of the powder fuel it is
reasonable to suppose that the free electron gas is formed near the surfaces
of grains of powder. But if a free electron gas interacts with the $%
LiAlH_{4}-Ni$ powder mixture applied then the above observations can be well
explained by the electron assisted neutron exchange processes. $_{3}^{7}Li$
has $\Delta _{-}=0.8214$ $MeV$ so it is able to lose neutron. The $\Delta
_{+}$values of the $Ni$ isotopes can be found in Table I. Completing Table I
with $\Delta _{+}(_{28}^{59}Ni)=3.319$ $MeV$ (the half life of $_{28}^{59}Ni$
is $\tau =7.6\times 10^{4}$ $y$) one can recognize that the $e+$ $%
_{3}^{7}Li+ $ $_{28}^{A}Ni\rightarrow e^{\prime }+$ $_{3}^{6}Li+$ $%
_{28}^{A+1}Ni+\Delta $ reaction has $\Delta >0$ value for $A=58-61$ but in
the case of $A=62$ the chain of reactions breaks since in this case $\Delta
<0$ because $\Delta _{+}(_{28}^{62}Ni)=-1.234$ $MeV$. The $64\rightarrow
63;61\rightarrow 62$ reaction of type $\left( \ref{NiAp}\right) $ (see Table
III) leads to production of $_{28}^{63}Ni$ ($\tau =100.1$ $y$) which has $%
\Delta _{-}(_{28}^{63}Ni)=1.2335$ $MeV$ allowing and coupling transition $%
63\rightarrow 62$ to transitions $58\rightarrow 59;59\rightarrow
60;60\rightarrow 61$ and $61\rightarrow 62$ in reaction $\left( \ref{NiAp}%
\right) $. These facts explain the enrichment of $_{28}^{62}Ni$ and $%
_{3}^{6}Li$ and the depletion of $_{3}^{7}Li$ and $Ni$ isotopes of $A=58-61$
and $64$. (Reactions $\left( \ref{NiAp}\right) $ too contribute to the
enrichment of $_{28}^{62}Ni$ (see Table III).)

\section{Conclusion}

It is thought that, in principle, the electron assisted processes are able
to answer the questions raised in the introduction. The exchange of the
original, extremely small Coulomb factor to the Coulomb factor of order of
unity of the electron in electron assisted processes answers problem (a).
The electron assisted nuclear reactions and the reactions which are coupled
with them are not accompanied by the expected nuclear end products answering
problem (b). Problem (c), the asserted appearance of nuclear transmutations
is partly answered in Section IV.C. with the aid of charged particle
assisted and usual nuclear reactions.

Summarizing, the theoretical results expounded and their successful
applications in explaining some unresolved experimental facts inspire us to
say that the studying of charged particles electron assisted nuclear
reactions, especially the electron assisted neutron exchange processes may
start a renaissance in the field of low energy nuclear physics.

\section{Appendix}

\subsection{Initial, intermediate and final states of electron assisted
neutron exchange process}

Let $\Psi _{i}$, $\Psi _{\mu }$ and $\Psi _{f}$ denote the space dependent
parts of initial, intermediate and final states, respectively. The initial
state has the form 
\begin{equation}
\Psi _{i}(\mathbf{x}_{e},\mathbf{x}_{1},\mathbf{x}_{n1},\mathbf{x}_{2})=\psi
_{ie}\left( \mathbf{x}_{e}\right) \psi _{i1n}(\mathbf{x}_{1},\mathbf{x}%
_{n1})\psi _{i2}(\mathbf{x}_{2}),  \label{Pszii}
\end{equation}%
where 
\begin{equation}
\psi _{ie}\left( \mathbf{x}_{e}\right) =V^{-1/2}e^{\left( i\mathbf{k}%
_{ie}\cdot \mathbf{x}_{e}\right) }\text{ and }\psi _{i2}(\mathbf{x}%
_{2})=V^{-1/2}e^{\left( i\mathbf{k}_{i2}\cdot \mathbf{x}_{2}\right) }
\label{psziei}
\end{equation}%
are the initial state of the electron and the nucleus $_{Z}^{A_{2}}X$, and $%
\psi _{i1n}(\mathbf{x}_{1},\mathbf{x}_{n1})$ is the initial state of the
neutron and the initial $A_{1}-1$ nucleon of the nucleus $_{Z}^{A_{1}}X$. $%
\mathbf{x}_{e}$, $\mathbf{x}_{1},\mathbf{x}_{n1}$ and $\mathbf{x}_{2}$ are
the coordinates of the electron, the center of mass of the initial $A_{1}-1$
nucleon, the neutron and the nucleus $_{Z}^{A_{2}}X$, respectively. $\mathbf{%
k}_{ie}$ and $\mathbf{k}_{i2}$ are the initial wave vectors of the electron
and the nucleus $_{Z}^{A_{2}}X$ and $V$ is the volume of normalization. The
initial state $\psi _{i1n}(\mathbf{x}_{1},\mathbf{x}_{n1})$ of the neutron
and the initial $A_{1}-1$ nucleon may be given in the variables $\mathbf{R}%
_{1}$, $\mathbf{r}_{n1}$ 
\begin{equation}
\psi _{i1n}(\mathbf{R}_{1},\mathbf{r}_{n1})=V^{-1/2}\exp (i\mathbf{k}%
_{i1}\cdot \mathbf{R}_{1})\Phi _{i1}\left( \mathbf{r}_{n1}\right)
\label{pszii1}
\end{equation}%
where $\mathbf{R}_{1}$ is the center of mass coordinate of the nucleus $%
_{Z}^{A_{1}}X$ and $\mathbf{r}_{n1}$ is the relative coordinate of one of
its neutrons. $\mathbf{R}_{1}$ and $\mathbf{r}_{n1}$are determined by the
usual $\mathbf{x}_{n1}=\mathbf{R}_{1}+\mathbf{r}_{n1}$ and $\mathbf{R}_{1}=%
\left[ \left( A_{1}-1\right) \mathbf{x}_{1}+\mathbf{x}_{n1}\right] /A_{1}$
relations where $\mathbf{x}_{n1}$ and $\mathbf{x}_{1}$ are the coordinates
of the neutron and of the center of mass of the initial $A_{1}-1$ nucleon,
respectively. The inverse formula for $\mathbf{x}_{1}$ is $\mathbf{x}_{1}=%
\mathbf{R}_{1}-\mathbf{r}_{n1}/\left( A_{1}-1\right) $. In $\left( \ref%
{pszii1}\right) $ the $\Phi _{i1}\left( \mathbf{r}_{n1}\right) $ is the wave
function of the neutron in the initial bound state of nucleus $_{Z}^{A_{1}}X$%
, $\mathbf{k}_{i1}$is the initial wave vector of nucleus $_{Z}^{A_{1}}X$.

The intermediate state has the form 
\begin{equation}
\Psi _{\mu }(\mathbf{x}_{e},\mathbf{x}_{1},\mathbf{x}_{n1},\mathbf{x}%
_{2})=\psi _{fe}\left( \mathbf{x}_{e}\right) \psi _{\mu 1n}(\mathbf{x}_{1},%
\mathbf{x}_{n1})\psi _{i2}(\mathbf{x}_{2}),  \label{Pszimu}
\end{equation}%
where 
\begin{equation}
\psi _{fe}\left( \mathbf{x}_{e}\right) =V^{-1/2}e^{\left( i\mathbf{k}%
_{fe}\cdot \mathbf{x}_{e}\right) }  \label{pszief}
\end{equation}%
with $\mathbf{k}_{fe}$ the wave vector of the electron in the final state
and $\psi _{i2}(\mathbf{x}_{2})$ is given in $\left( \ref{psziei}\right) $.
The state $\psi _{\mu 1n}(\mathbf{x}_{1},\mathbf{x}_{n1})$ is the product of
two plane waves $\psi _{f1}(\mathbf{x}_{1})=V^{-1/2}e^{\left( i\mathbf{k}%
_{1}\cdot \mathbf{x}_{1}\right) }$ and $\psi _{n}\left( \mathbf{x}%
_{n1}\right) =V^{-1/2}e^{i\mathbf{k}_{n}\cdot \mathbf{x}_{n1}}$, which are
the final state of the nucleus $_{Z_{1}}^{A_{1}-1}X$ and the state of the
free, intermediate neutron. Thus $\psi _{\mu 1n}(\mathbf{x}_{1},\mathbf{x}%
_{n1})=V^{-1}e^{i\mathbf{k}_{1}\cdot \mathbf{x}_{1}}e^{i\mathbf{k}_{n}\cdot 
\mathbf{x}_{n1}}$ and it has the form in the coordinates $\mathbf{R}_{1}$, $%
\mathbf{r}_{n1}$%
\begin{equation}
\psi _{\mu 1n}(\mathbf{R}_{1},\mathbf{r}_{n1})=V^{-1}e^{i\left( \mathbf{k}%
_{1}+\mathbf{k}_{n}\right) \cdot \mathbf{R}_{1}}e^{i\left( \mathbf{k}_{n}-%
\frac{\mathbf{k}_{1}}{A_{1}-1}\right) \mathbf{r}_{n1}},  \label{pszimu2}
\end{equation}%
where $\mathbf{k}_{1}$ and $\mathbf{k}_{n}$ are the wave vectors of the
nucleus $_{Z}^{A_{1}-1}X$ and the neutron, respectively.

The intermediate state may have an other form%
\begin{equation}
\Psi _{\mu }(\mathbf{x}_{e},\mathbf{x}_{1},\mathbf{x}_{n1},\mathbf{x}%
_{2})=\psi _{fe}\left( \mathbf{x}_{e}\right) \psi _{f1}(\mathbf{x}_{1})\psi
_{\mu 2n}(\mathbf{x}_{n1},\mathbf{x}_{2}),  \label{Pszimu2}
\end{equation}%
where 
\begin{equation}
\psi _{\mu 2n}(\mathbf{x}_{n1},\mathbf{x}_{2})=\psi _{n}\left( \mathbf{x}%
_{n1}\right) \psi _{i2}(\mathbf{x}_{2})=V^{-1}e^{i\mathbf{k}_{n}\cdot 
\mathbf{x}_{n1}}e^{i\mathbf{k}_{i2}\cdot \mathbf{x}_{2}}  \label{pszimu3}
\end{equation}%
which can be written in the coordinates $\mathbf{r}_{n2}=\mathbf{x}_{n1}-%
\mathbf{R}_{2}$ and $\mathbf{R}_{2}=\left( A_{2}\mathbf{x}_{2}+\mathbf{x}%
_{n1}\right) /\left( A_{2}+1\right) $ as 
\begin{equation}
\psi _{\mu 2n}(\mathbf{R}_{2},\mathbf{r}_{n2})=\frac{1}{V}e^{i\left( \mathbf{%
k}_{i2}+\mathbf{k}_{n}\right) \cdot \mathbf{R}_{2}}e^{i\left( \mathbf{k}_{n}-%
\frac{\mathbf{k}_{i2}}{A_{2}}\right) \mathbf{r}_{n2}},  \label{pszimu4}
\end{equation}%
where $\mathbf{R}_{2}$ is the center of mass coordinate of the nucleus $%
_{Z}^{A_{2}+1}X$ and $\mathbf{r}_{n2}$ is the relative coordinate of the
neutron in it. In these new variables $\mathbf{x}_{2}=\mathbf{R}_{2}-\mathbf{%
r}_{n2}/A_{2}$ and $\mathbf{x}_{n1}-\mathbf{x}_{2}=\left( A_{2}+1\right) 
\mathbf{r}_{n2}/A_{2}$ which is used in the argument of $V^{St}$ (given by $%
\left( \ref{VSt1}\right) $) in calculating $V_{f\mu }^{St}$. Evaluating the
matrix elements $V_{\mu i}^{Cb}$ and $V_{f\mu }^{St}$ the forms $\left( \ref%
{pszimu2}\right) $ and $\left( \ref{pszimu4}\right) $ of $\psi _{\mu }$ are
used, respectively, and $\sum_{\mu }\rightarrow \frac{V}{\left( 2\pi \right)
^{3}}d^{3}k_{n}$ in $\left( \ref{Tif}\right) $.

The final state has the form 
\begin{equation}
\Psi _{f}(\mathbf{x}_{e},\mathbf{x}_{1},\mathbf{x}_{n1},\mathbf{x}_{2})=\psi
_{fe}\left( \mathbf{x}_{e}\right) \psi _{f1}(\mathbf{x}_{1})\psi _{f2n}(%
\mathbf{x}_{n1},\mathbf{x}_{2}),  \label{Pszif}
\end{equation}%
where $\psi _{f2n}(\mathbf{x}_{n1},\mathbf{x}_{2})$ is given in the
variables $\mathbf{R}_{2}$, $\mathbf{r}_{n2}$ as 
\begin{equation}
\psi _{f2n}(\mathbf{R}_{2},\mathbf{r}_{n2})=V^{-1/2}\exp (i\mathbf{k}%
_{2}\cdot \mathbf{R}_{2})\Phi _{f2}\left( \mathbf{r}_{n2}\right) ,
\label{pszif2}
\end{equation}%
and $\Phi _{f2}\left( \mathbf{r}_{n2}\right) $ is the bound state of the
neutron in the nucleus $_{Z}^{A_{2}+1}X$.

\subsection{Evaluation of matrix elements $V_{\protect\mu i}^{Cb}$ and $V_{f%
\protect\mu }^{St}$}

The argument of the Coulomb potential $V^{Cb}$ is $\mathbf{x}_{e}-\mathbf{x}%
_{1}$ therefore the integration with respect to the components of $\mathbf{x}%
_{2}$ may be carried out and $\int \left\vert \psi _{i2}(\mathbf{x}%
_{2})\right\vert ^{2}d^{3}x_{2}=1$. The remainder is 
\begin{eqnarray}
V_{\mu i}^{Cb} &=&\int \psi _{fe}^{\ast }\left( \mathbf{x}_{e}\right) \psi
_{\mu 1n}^{\ast }(\mathbf{x}_{1},\mathbf{x}_{n1})V^{Cb}\left( \mathbf{x}_{e}-%
\mathbf{x}_{1}\right)  \label{VCbmui} \\
&&\times \psi _{ie}\left( \mathbf{x}_{e}\right) \psi _{i1n}(\mathbf{x}_{1},%
\mathbf{x}_{n1})d^{3}x_{e}d^{3}x_{1}d^{3}x_{n1}.  \notag
\end{eqnarray}%
Making the $\mathbf{x}_{1},\mathbf{x}_{n1}\rightarrow \mathbf{R}_{1},\mathbf{%
r}_{n1}$ change in the variables, substituting the forms $\left( \ref{pszii1}%
\right) $ and $\left( \ref{pszimu2}\right) $ of $\psi _{i1n}$ and $\psi
_{\mu 1n}$, and neglecting $\mathbf{k}_{i1}$, the integrations over the
components of $\mathbf{x}_{e}$ and $\mathbf{R}_{1}$ result $V^{-1}\left(
2\pi \right) ^{3}\delta \left( \mathbf{q}+\mathbf{k}_{ie}-\mathbf{k}%
_{fe}\right) $ and $V^{-3/2}\left( 2\pi \right) ^{3}\delta \left( \mathbf{q}-%
\mathbf{k}_{1}-\mathbf{k}_{n}\right) $, respectively and the integration
over the components of $\mathbf{r}_{n1}$ produces $F_{1}\left( \mathbf{k}%
_{n}\right) $ where 
\begin{equation}
F_{1}\left( \mathbf{k}_{n}\right) =\int \Phi _{i1}\left( \mathbf{r}%
_{n1}\right) e^{-i\left( \mathbf{k}_{n}-\frac{\mathbf{k}_{1}+\mathbf{q}}{%
A_{1}-1}\right) \cdot \mathbf{r}_{n1}}d^{3}r_{n1}.  \label{Fk1kn}
\end{equation}%
Using the $\delta \left( \mathbf{q}+\mathbf{k}_{ie}-\mathbf{k}_{fe}\right) $
in carrying out the integration over the components of $\mathbf{q}$ in $%
V_{\mu i}^{Cb}$ one gets%
\begin{eqnarray}
V_{\mu i}^{Cb} &=&-\frac{Ze^{2}}{2\pi ^{2}\left\vert \mathbf{k}_{fe}-\mathbf{%
k}_{ie}\right\vert ^{2}+\lambda ^{2}}\widetilde{F}_{1}\left( \mathbf{k}%
_{n}\right) \frac{\left( 2\pi \right) ^{6}}{V^{5/2}}\times  \label{VCbmui1}
\\
&&\times \sqrt{G_{S}}\delta \left( \mathbf{k}_{ie}-\mathbf{k}_{fe}-\mathbf{k}%
_{1}-\mathbf{k}_{n}\right)  \notag
\end{eqnarray}%
and%
\begin{equation}
\widetilde{F}_{1}\left( \mathbf{k}_{n}\right) =\int \Phi _{i1}\left( \mathbf{%
r}_{n1}\right) e^{-i\left( \mathbf{k}_{n}-\frac{\mathbf{k}_{1}+\mathbf{k}%
_{fe}-\mathbf{k}_{ie}}{A_{1}-1}\right) \cdot \mathbf{r}_{n1}}d^{3}r_{n1}.
\label{Fk1kn2}
\end{equation}%
For particles $e$ and $1$ (ingoing electron of charge $-e$ and initial
nucleus $_{Z}^{A_{1}}X$ of charge $Ze$) taking part in Coulomb interaction
we have used plane waves therefore the matrix element must be corrected with
the so called Sommerfeld factor \cite{Heitler} $\sqrt{G_{S}}$ where%
\begin{equation}
G_{S}=\frac{F_{e}(E_{ie})}{F_{e}(E_{f1})}.  \label{Gs}
\end{equation}

Now we deal with $V_{f\mu }^{St}$. The strong interaction works between the
neutron and the nucleons of the nucleus $_{Z}^{A_{2}}X$ therefore the
argument of $V^{St}$ is $\mathbf{x}_{n1}-\mathbf{x}_{2}$. The integrations
with respect to the components of $\ \mathbf{x}_{e}$ and $\mathbf{x}_{1}$
result $\int \left\vert \psi _{ef}(\mathbf{x}_{e})\right\vert
^{2}d^{3}x_{e}= $ $\int \left\vert \psi _{f1}(\mathbf{x}_{1})\right\vert
^{2}d^{3}x_{1}=1$. The remainder is 
\begin{equation}
V_{f\mu }^{St}=\int \psi _{f2n}^{\ast }V^{St}\left( \mathbf{x}_{n1}-\mathbf{x%
}_{2}\right) \psi _{\mu 2n}d^{3}x_{2}d^{3}x_{n1}.  \label{VStfmu}
\end{equation}%
Similarly to the above, making the $\mathbf{x}_{n1},\mathbf{x}%
_{2}\rightarrow \mathbf{R}_{2},\mathbf{r}_{n2}$ change in the variables,
substituting the forms $\left( \ref{pszimu4}\right) $ and $\left( \ref%
{pszif2}\right) $ of $\psi _{\mu 2n}$ and $\psi _{f2n}^{\ast }$ and
neglecting $\mathbf{k}_{i2}$, the integrations over the components of $%
\mathbf{R}_{2}$ result $V^{-3/2}\left( 2\pi \right) ^{3}\delta \left( 
\mathbf{k}_{n}-\mathbf{k}_{2}\right) $ and the integrations with respect to
the components of $\mathbf{r}_{n2}$ produces $F_{2}\left( \mathbf{k}%
_{n}\right) $ with%
\begin{eqnarray}
F_{2}\left( \mathbf{k}_{n}\right) &=&\int \Phi _{f2}^{\ast }\left( \mathbf{r}%
_{n2}\right) e^{i\mathbf{k}_{n}\cdot \mathbf{r}_{n2}}\times  \label{F2ki2kn}
\\
&&\times \left( -f\frac{\exp (-s\frac{A_{2}+1}{A_{2}}r_{n2}}{\frac{A_{2}+1}{%
A_{2}}r_{n2}}\right) d^{3}r_{n2},  \notag
\end{eqnarray}%
where $r_{n2}=\left\vert \mathbf{r}_{n2}\right\vert $. Taking into account
that the neutron interacts with each nucleon of the final nucleus of nucleon
number $A_{2}$ 
\begin{equation}
V_{f\mu }^{St}=\frac{\left( 2\pi \right) ^{3}}{V^{3/2}}A_{2}F_{2}\left( 
\mathbf{k}_{n}\right) \delta \left( \mathbf{k}_{n}-\mathbf{k}_{2}\right) .
\label{VStfmu2}
\end{equation}

\subsection{Transition probability per unit time of electron assisted
neutron exchange process}

Substituting the obtained forms of $V_{\mu i}^{Cb}$ and $V_{f\mu }^{St}$
(formulae $\left( \ref{VCbmui1}\right) $ and $\left( \ref{VStfmu2}\right) $)
into $\left( \ref{Tif}\right) $ and using the correspondence $\sum_{\mu
}\rightarrow \frac{V}{\left( 2\pi \right) ^{3}}d^{3}k_{n}$ and the $\delta
\left( \mathbf{k}_{n}-\mathbf{k}_{2}\right) $ in the integration over the
components of $\mathbf{k}_{n}$ one gets%
\begin{eqnarray}
T_{fi} &=&-\frac{e^{2}ZA_{2}\widetilde{F}_{1}\left( \mathbf{k}_{2}\right)
F_{2}\left( \mathbf{k}_{2}\right) \sqrt{\frac{F_{e}(E_{ie})}{F_{e}(E_{f1})}}%
}{2\pi ^{2}\left\vert \mathbf{k}_{fe}-\mathbf{k}_{ie}\right\vert
^{2}+\lambda ^{2}}\times  \label{Tfi22} \\
&&\times \frac{\left( 2\pi \right) ^{6}}{V^{3}}\frac{\delta \left( \mathbf{k}%
_{1}+\mathbf{k}_{2}+\mathbf{k}_{fe}-\mathbf{k}_{ie}\right) }{\left( \Delta
E_{\mu i}\right) _{\mathbf{k}_{n}=\mathbf{k}_{2}}},  \notag
\end{eqnarray}%
where%
\begin{equation}
\widetilde{F}_{1}\left( \mathbf{k}_{2}\right) =\int \Phi _{i1}\left( \mathbf{%
r}_{n1}\right) e^{-i\left( \mathbf{k}_{2}-\frac{\mathbf{k}_{1}+\mathbf{k}%
_{fe}-\mathbf{k}_{ie}}{A_{1}-1}\right) \cdot \mathbf{r}_{n1}}d^{3}r_{n1}
\label{F1k2}
\end{equation}%
and $F_{2}\left( \mathbf{k}_{2}\right) $ is determined by $\left( \ref{F2k2}%
\right) $. Here $\Phi _{i1}$ and $\Phi _{f2}$ in $\left( \ref{F2k2}\right) $
are the initial and final bound neutron states. Substituting the above into $%
\left( \ref{Wfie}\right) $, using the identities $\left[ \delta \left( 
\mathbf{k}_{1}+\mathbf{k}_{2}+\mathbf{k}_{fe}-\mathbf{k}_{ie}\right) \right]
^{2}=\delta \left( \mathbf{k}_{1}+\mathbf{k}_{2}+\mathbf{k}_{fe}-\mathbf{k}%
_{ie}\right) \delta \left( \mathbf{0}\right) $ and $\left( 2\pi \right)
^{3}\delta \left( \mathbf{0}\right) =V$, the $\sum_{f}\rightarrow
\sum_{m_{2}}\int \left[ V/\left( 2\pi \right) ^{3}\right]
^{3}d^{3}k_{1}d^{3}k_{2}d^{3}k_{fe}$ correspondence, averaging over the
quantum number $m_{1\text{ }}$and integrating over the components of $%
\mathbf{k}_{fe}$ (which gives $\mathbf{k}_{fe}=-\mathbf{k}_{1}-\mathbf{k}%
_{2}+\mathbf{k}_{ie}$) one obtains%
\begin{eqnarray}
W_{fi} &=&\int \frac{\alpha _{f}^{2}\hbar
c^{2}Z^{2}\sum_{l_{2}=-m_{2}}^{l_{2}=m_{2}}\left\vert F_{2}\left( \mathbf{k}%
_{2}\right) \right\vert ^{2}}{\pi ^{3}v_{c}V\left( \left\vert \mathbf{k}_{1}+%
\mathbf{k}_{2}\right\vert ^{2}+\lambda ^{2}\right) ^{2}\left( \Delta E_{\mu
i}\right) _{\mathbf{k}_{n}=\mathbf{k}_{2}}^{2}}  \label{Wfi22} \\
&&\times \left\langle \left\vert F_{1}\left( \mathbf{k}_{2}\right)
\right\vert ^{2}\right\rangle \frac{F_{e}(E_{ie})}{F_{e}(E_{f1})}%
A_{2}^{2}r_{A_{2}}\delta (E_{f}-\Delta )d^{3}k_{1}d^{3}k_{2},  \notag
\end{eqnarray}%
where $A_{1}$, $A_{2}$ are the initial atomic masses, $l_{1},m_{1}$ and $%
l_{2},m_{2}$ are the orbit and its projection quantum numbers of the neutron
in its initial and final state. For $F_{1}\left( \mathbf{k}_{2}\right) $, $%
\left\langle \left\vert F_{1}\left( \mathbf{k}_{2}\right) \right\vert
^{2}\right\rangle $ and $F_{2}\left( \mathbf{k}_{2}\right) $ see $\left( \ref%
{F1kalk2}\right) $, $\left( \ref{F1av}\right) $ and $\left( \ref{F2k2}%
\right) $. Taking into account the effect of the number of atoms of atomic
number $A_{2}$ in the solid target the calculation is similar to the
calculation of e.g. the coherent neutron scattering \cite{Kittel} and the $%
\left\vert T_{fi}\right\vert ^{2}$ must be multiplied by $N_{L}$ which is
the number of atomic sites in the crystal and by $r_{A_{2}}$ which is the
relative natural abundance of atoms $_{Z}^{A_{2}}X$. We have used $%
N_{L}/V=2/v_{c}$ with $v_{c}$ the volume of the elementary cell of the $fcc$
lattice in which there are two lattice sites in the cases of $Ni$ and $Pd$
investigated.

\subsection{Approximations, identities and relations in calculation of cross
section}

Now we deal with the energy denominator $\left( \Delta E_{\mu i}\right) $ in 
$\left( \ref{Wfi22}\right) $ and $\left( \ref{sigma}\right) $ $\left[ \text{%
see }\left( \ref{DeltaEmui}\right) -\left( \ref{E1f}\right) \right] $. The
shielding parameter $\lambda $ is determined by the innermost electronic
shell of the atom $_{Z}^{A_{1}}X$ and it can be determined as%
\begin{equation}
\lambda =\frac{Z}{a_{B}},  \label{lambda}
\end{equation}%
where $a_{B}=0.53\times 10^{-8}$ $cm$ is the Bohr-radius. The integrals in $%
\left( \ref{Wfi22}\right) $ and $\left( \ref{sigma}\right) $ have
accountable contributions if%
\begin{equation}
\ \left\vert \mathbf{k}_{1}+\mathbf{k}_{2}\right\vert \lesssim \lambda
\label{condlambda}
\end{equation}%
and then $E_{fe}\lesssim \hbar ^{2}\lambda ^{2}/\left( 2m_{e}\right) =\frac{1%
}{2}\alpha _{f}^{2}m_{e}c^{2}Z^{2}$ which can be neglected in $\Delta E_{\mu
i}$ and in the energy Dirac-delta. Thus%
\begin{equation}
\Delta E_{\mu i}=\frac{\hbar ^{2}\mathbf{k}_{1}^{2}}{2m_{1}}+\frac{\hbar ^{2}%
\mathbf{k}_{2}^{2}}{2m_{n}}-\Delta _{-}+\Delta _{n}  \label{DeltaEmui2}
\end{equation}%
and in the Dirac-delta%
\begin{equation}
E_{f}=\frac{\hbar ^{2}\mathbf{k}_{1}^{2}}{2m_{1}}+\frac{\hbar ^{2}\mathbf{k}%
_{2}^{2}}{2m_{2}}.  \label{Ef2}
\end{equation}%
In this case $\mathbf{k}_{1}=-\mathbf{k}_{2}+\delta \mathbf{k}$ with $%
\left\vert \delta \mathbf{k}\right\vert =\delta k\sim \lambda $. Using 
\begin{equation}
k_{1}\simeq k_{2}\simeq k_{0}=\sqrt{2\mu _{12}\Delta }/\hbar  \label{k0}
\end{equation}%
(see below) with $\mu _{12}c^{2}=A_{12}m_{0}c^{2}$, where $A_{12}=\left(
A_{1}-1\right) \left( A_{2}+1\right) /\left( A_{1}+A_{2}\right) $ is the
reduced nucleon number, one can conclude that the $\mathbf{k}_{2}=-\mathbf{k}%
_{1}$ relation fails with a very small error in the cases of events which
fulfill condition $\left( \ref{condlambda}\right) $ since $k_{1}/k_{0}\simeq
1$, $k_{2}/k_{0}\simeq 1$,$\ \delta k/k_{0}\sim \lambda /k_{0}$ and $\lambda
/k_{0}=\alpha _{f}Zm_{e}c^{2}/\sqrt{2\mu _{12}c^{2}\Delta }\ll 1$.
Consequently, the quantity $E_{f}$ in the argument of the energy Dirac-delta
can be written approximately as 
\begin{equation}
E_{f}=\left( \frac{\hbar ^{2}}{2m_{1}}+\frac{\hbar ^{2}}{2m_{2}}\right) 
\mathbf{k}_{2}^{2}=\frac{\hbar ^{2}c^{2}\mathbf{k}_{2}^{2}}{2A_{12}m_{0}c^{2}%
}.  \label{Ef22}
\end{equation}%
Furthermore taking $A_{1}/\left( A_{1}+1\right) \simeq 1$ 
\begin{equation}
\Delta E_{\mu i}=\frac{\hbar ^{2}c^{2}\mathbf{k}_{2}^{2}}{2m_{0}c^{2}}%
-\Delta _{-}+\Delta _{n}.  \label{DeltaEmui3}
\end{equation}

We introduce the $\mathbf{Q}=\hbar c\mathbf{k}_{2}/\Delta $, $\mathbf{P}%
=\hbar c\left( \delta \mathbf{k}\right) /\Delta $, $\varepsilon
_{f}=E_{f}/\Delta =\left[ \mathbf{Q}^{2}/\left( 2A_{12}m_{0}c^{2}\right) %
\right] \Delta $ and $L=\hbar c\lambda /\Delta $ dimensionless quantities.
The energy Dirac-delta modifies as $\delta (E_{f}-\Delta )=\delta \left[
\varepsilon _{f}\left( \mathbf{Q}\right) -1\right] /\Delta $. \ The relation 
$\left( \ref{lambda}\right) $ yields $L=\hbar cZ/\left( a_{B}\Delta \right)
=Z\alpha _{f}m_{e}c^{2}/\Delta $ and $Z\alpha _{f}m_{e}c^{2}/\Delta \lesssim
1$. Now we change $d^{3}k_{1}d^{3}k_{2}$ to $\left( \frac{\Delta }{\hbar c}%
\right) ^{6}d^{3}Qd^{3}P$ in the integration in $\left( \ref{sigma}\right) $%
, use the $\delta \left[ g\left( Q\right) \right] =\delta \left(
Q-Q_{0}\right) /g^{\prime }\left( Q_{0}\right) $ identity, where $Q_{0}$ is
the root of the equation $g\left( Q\right) =0$ ($k_{0}=Q_{0}\Delta /\left(
\hbar c\right) $, see $\left( \ref{k0}\right) $), estimate the integral with
respect to the components of $\mathbf{P}$ by 
\begin{equation}
\int_{0}^{\infty }\frac{4\pi P^{2}dP}{\left( P^{2}+L^{2}\right) ^{2}}=\frac{%
\pi ^{2}}{L}  \label{IntP}
\end{equation}%
and apply $v_{c}=d^{3}/4$ (the volume of unit cell of $fcc$ lattice for $Ni$
and $Pd$ of lattice parameter $d$).

\subsection{$\left\langle \left\vert F_{1}\left( \mathbf{k}_{0}\right)
\right\vert ^{2}\right\rangle _{Sh}$ and $\sum_{l_{2}=-m_{2}}^{l_{2}=m_{2}}%
\left\vert F_{2}\left( \mathbf{k}_{0}\right) \right\vert _{Sh}^{2}$\ in
single particle shell-model and without LWA}

Now we calculate the quantities $\left\langle \left\vert F_{1}\left( \mathbf{%
k}_{0}\right) \right\vert ^{2}\right\rangle _{Sh}$~and $%
\sum_{l_{2}=-m_{2}}^{l_{2}=m_{2}}\left\vert F_{2}\left( \mathbf{k}%
_{0}\right) \right\vert _{Sh}^{2}$ in the single particle shell model with
isotropic harmonic oscillator potential and without the long wavelength
approximation (see definitions: $\left( \ref{F1kalk2}\right) $, $\left( \ref%
{F1av}\right) $ and $\left( \ref{F2k2}\right) $). Taking into account the
spin-orbit coupling in the level scheme the emerging neutron states are $0l$
and $1l$ shell model states in the cases of $Ni$ and $Pd$ to be discussed
numerically \cite{Pal}. So the initial and final neutron states $\left( \Phi
_{i1},\Phi _{f2}\right) $ have the form%
\begin{equation}
\Phi _{Sh}\left( \mathbf{r}_{nj}\right) =\frac{R_{n_{j}l_{j}}}{r_{nj}}%
Y_{l_{j}m_{j}}\left( \Omega _{j}\right)  \label{Fishell}
\end{equation}%
where $n_{j}=0,1$ in the cases of $0l$ and $1l$ investigated, respectively,
and 
\begin{equation}
R_{0l_{j}}=b_{j}^{-1/2}\left( \frac{2}{\Gamma (l_{j}+3/2)}\right)
^{1/2}\varrho _{j}^{l_{j}+1}\exp \left( -\frac{1}{2}\varrho _{j}^{2}\right) ,
\label{R0l}
\end{equation}%
\begin{eqnarray}
R_{1l_{j}} &=&b_{j}^{-1/2}\left( \frac{2l_{j}+3}{\Gamma (l_{j}+3/2)}\right)
^{1/2}\varrho _{j}^{l_{j}+1}\times  \label{R1l} \\
&&\times \left( 1-\frac{2}{2l_{j}+3}\varrho _{j}^{2}\right) \exp \left( -%
\frac{1}{2}\varrho _{j}^{2}\right)  \notag
\end{eqnarray}%
with $\varrho _{j}=r_{nj}/b_{j}$ where $b_{j}=\sqrt{\hbar /\left(
m_{0}\omega _{j}\right) }$ \cite{Pal}. Here $\omega _{j}$ is the angular
frequency of the oscillator that is determined by $\hbar \omega
_{1}=40A_{1}^{-1/3}$ $MeV$ and\ $\hbar \omega _{2}=40\left( A_{2}+1\right)
^{-1/3}$ $MeV$ \cite{Bohr}. (The subscript $Sh$ refers to the shell model.)
With the aid of these wave functions and for $n_{1}=0,1$ 
\begin{equation}
\left\langle \left\vert F_{1}\left( \mathbf{k}_{0}\right) \right\vert
^{2}\right\rangle _{Sh}=b_{1}^{3}\frac{2^{l_{1}+2}}{\sqrt{\pi }\left(
2l_{1}+1\right) !!}4\pi I_{1,n_{1}}^{2}  \label{F1K2Sh}
\end{equation}%
with 
\begin{equation}
I_{1,0}=\int_{0}^{\infty }\varrho ^{l_{1}+2}j_{l_{1}}(k_{0}b_{1}\frac{A_{1}}{%
A_{1}-1}\varrho )e^{-\frac{1}{2}\varrho ^{2}}d\varrho \text{ }  \label{I10}
\end{equation}%
and 
\begin{eqnarray}
I_{1,1} &=&\left( l_{1}+\frac{3}{2}\right) \int_{0}^{\infty }\varrho
^{l_{1}+2}\left( 1-\frac{2}{2l_{1}+3}\varrho ^{2}\right) \times  \label{I11}
\\
&&\times j_{l_{1}}(k_{0}b_{1}\frac{A_{1}}{A_{1}-1}\varrho )e^{-\frac{1}{2}%
\varrho ^{2}}d\varrho .\text{ }  \notag
\end{eqnarray}%
Here $j_{l_{1}}(x)=\sqrt{\frac{\pi }{2x}}J_{l_{1}+1/2}(x)$ denotes spherical
Bessel function with $J_{l_{1}+1/2}(x)$ the Bessel function of first kind.

Similarly%
\begin{eqnarray}
\sum_{l_{2}=-m_{2}}^{l_{2}=m_{2}}\left\vert F_{2}\left( \mathbf{k}%
_{0}\right) \right\vert _{Sh}^{2} &=&b_{2}f^{2}\frac{2^{l_{2}+2}\left(
2l_{2}+1\right) }{\sqrt{\pi }\left( 2l_{2}+1\right) !!}\times  \label{F2k2Sh}
\\
&&\times 4\pi \left( \frac{A_{2}}{A_{2}+1}\right) ^{2}I_{2,n_{2}}^{2}  \notag
\end{eqnarray}%
with%
\begin{equation}
I_{2,0}=\int_{0}^{\infty }\varrho ^{l_{2}+1}j_{l_{2}}(k_{0}b_{2}\varrho )e^{-%
\frac{1}{2}\varrho ^{2}-\frac{A_{2}+1}{A_{2}}\frac{b_{2}}{r_{0}}\varrho
}d\varrho \text{ }  \label{I20}
\end{equation}%
and%
\begin{eqnarray}
I_{2,1} &=&\left( l_{2}+\frac{3}{2}\right) \int_{0}^{\infty }\varrho
^{l_{2}+1}\left( 1-\frac{2}{2l_{2}+3}\varrho ^{2}\right) \times  \label{I21}
\\
&&\times j_{l_{2}}(k_{0}b_{2}\varrho )e^{-\frac{1}{2}\varrho ^{2}-\frac{%
A_{2}+1}{A_{2}}\frac{b_{2}}{r_{0}}\varrho }d\varrho .  \notag
\end{eqnarray}%
Substituting the results of $\left( \ref{F1K2Sh}\right) $, $\left( \ref%
{F2k2Sh}\right) $ and $\left( \ref{F1K2av}\right) $ into $\left( \ref{etha}%
\right) $ one gets%
\begin{eqnarray}
\eta _{l_{1},n_{1},l_{2},n_{2}}\left( A_{1},A_{2}\right) &=&\frac{%
2^{l_{1}+l_{2}+4}}{\pi \left( 2l_{1}+1\right) !!\left( 2l_{2}+1\right) !!}%
\times  \label{etha2} \\
&&\times \frac{b_{1}^{3}b_{2}}{r_{0}^{4}}\left( \frac{A_{2}}{A_{2}+1}\right)
^{2}I_{1,n_{1}}^{2}I_{2,n_{2}}^{2}.  \notag
\end{eqnarray}

\subsection{Coulomb factors $F_{2^{\prime }3}$ and $F_{34}$ in electron
assisted heavy charged particle exchange process}

If initial particles have negligible initial momentum then, because of
momentum conservation, $\mathbf{k}_{2^{\prime }}=-\mathbf{k}_{5}$ in the
final state. (It was obtained \cite{kk3} that the process has accountable
cross section if the momentum of the final electron can be neglected, i.e.
in the $\mathbf{k}_{1^{\prime }}\simeq 0$ case.) Thus the condition of
energy conservation 
\begin{equation}
\frac{\hbar ^{2}\mathbf{k}_{2^{\prime }}^{2}}{2m_{2^{\prime }}}+\frac{\hbar
^{2}\mathbf{k}_{5}^{2}}{2m_{5}}=\Delta  \label{ec}
\end{equation}%
determines $\mathbf{k}_{2^{\prime }}$ as%
\begin{equation}
\hbar ^{2}\mathbf{k}_{2^{\prime }}^{2}=2\mu _{2^{\prime }5}\Delta ,
\label{k2'}
\end{equation}%
where $\hbar $ is the reduced Planck-constant,%
\begin{equation}
\mu _{2^{\prime }5},=a_{2^{\prime }5}m_{0}c^{2}  \label{mu2'5}
\end{equation}%
is the reduced rest mass of particles $2^{\prime }$ and $5$ of mass numbers $%
A_{2^{\prime }}$ and $A_{5}$ [for $a_{2^{\prime }5}$ see $\left( \ref{ajk}%
\right) $]. If the initial momenta and the momentum of particle $1^{\prime }$
are negligible then $\mathbf{k}_{3}=-\mathbf{k}_{2^{\prime }}$, since
momentum is preserved in Coulomb scattering. Thus the energy $E_{3}$ of
particle $3$ can be written as 
\begin{equation}
E_{3}=\frac{\hbar ^{2}\mathbf{k}_{3}^{2}}{2m_{3}}=\frac{\mu _{2^{\prime }5}}{%
m_{3}}\Delta =\frac{a_{2^{\prime }5}}{A_{3}}\Delta .  \label{E3}
\end{equation}%
Calculating the Coulomb factor $F_{2^{\prime }3}$ [see $\left( \ref{Fjk}%
\right) $]\ between particles $2^{\prime }$ and $3$ the energy determined by 
$\left( \ref{E3}\right) $ is given in their $CM$ coordinate system (since $%
\mathbf{k}_{3}=-\mathbf{k}_{2^{\prime }}$) thus it can be substituted
directly in $\left( \ref{etajk}\right) $ producing%
\begin{equation}
\eta _{2^{\prime }3}=\left( Z_{2}-z_{3}\right) z_{3}\alpha _{f}A_{3}\sqrt{%
\frac{A_{2^{\prime }}+A_{5}}{\left( A_{2^{\prime }}+A_{3}\right) A_{5}}\frac{%
m_{0}c^{2}}{2\Delta }}.  \label{eta2'3}
\end{equation}%
Since the above analysis is made in order to discuss the phenomenon of
nuclear transmutation we take $A_{3}\ll A_{2^{\prime }}\simeq A_{5}=A$ ($%
\gtrsim 100$ in the case of $Pd$ discussed). So $\left( A_{2^{\prime
}}+A_{5}\right) /\left[ \left( A_{2^{\prime }}+A_{3}\right) A_{5}\right]
\simeq 2/A$ and $\eta _{2^{\prime }3}$ reads approximately as 
\begin{equation}
\eta _{2^{\prime }3}=\left( Z_{2}-z_{3}\right) z_{3}\alpha _{f}A_{3}\sqrt{%
\frac{m_{0}c^{2}}{A\Delta }}.  \label{eta2'3approx}
\end{equation}%
Calculating the Coulomb factor $F_{34}$, the energy of particle $3$
determined by $\left( \ref{E3}\right) $ is now given in the laboratory frame
of reference since particle $4$ is at rest. In the $CM$ system of particles $%
3$ and $4$ the energy $E_{3}(CM)$ is 
\begin{equation}
E_{3}(CM)=\frac{A_{4}a_{2^{\prime }5}\Delta }{\left( A_{3}+A_{4}\right) A_{3}%
}.  \label{E3CM}
\end{equation}%
Substituting it into $\left( \ref{etajk}\right) $ 
\begin{equation}
\eta _{34}=\left( Z_{4}+z_{3}\right) z_{3}\alpha _{f}A_{3}\sqrt{\frac{%
m_{0}c^{2}}{2a_{2^{\prime }5}\Delta }}.  \label{eta34}
\end{equation}%
Applying the same approximation as above in which $2a_{2^{\prime }5}\simeq A$
\begin{equation}
\eta _{34}=\left( Z_{4}+z_{3}\right) z_{3}\alpha _{f}A_{3}\sqrt{\frac{%
m_{0}c^{2}}{A\Delta }}.  \label{eta34approx}
\end{equation}%
Furthermore, if $Z_{2}\simeq Z_{4}=Z\gg z_{3}$ then%
\begin{equation}
\eta _{2^{\prime }3}=\eta _{34}=Zz_{3}\alpha _{f}A_{3}\sqrt{\frac{m_{0}c^{2}%
}{A\Delta },}  \label{F2'3F34}
\end{equation}%
consequently, $F_{2^{\prime }3}=F_{34}$.

\end{document}